\documentclass[10pt,a4paper,onecolumn]{article}
\usepackage[T1]{fontenc}
\usepackage{amssymb}
\usepackage[utf8]{inputenc}
\usepackage{setspace}
\usepackage[font=footnotesize]{caption}
\usepackage[biblabel]{cite}
\usepackage{graphicx}
\usepackage{float}
\usepackage{lineno}
\usepackage[dvipsnames]{xcolor}
\usepackage{placeins}
\usepackage{amsmath}
\usepackage{multicol}
\setpagewiselinenumbers
\let\oldbibliography\thebibliography
\renewcommand{\thebibliography}[1]{%
  \oldbibliography{#1}%
  \setlength{\itemsep}{0pt}%
}
\graphicspath{{Figs/}}

\begin{document}
\onehalfspacing
\pagenumbering{arabic}

\vspace*{\fill}

\begin{center}
{\Large{\bf {\Large{\bf Single photon emitters in monolayer semiconductors coupled to transition metal dichalcogenide nanoantennas on silica and gold substrates}}}}
\vskip1.0\baselineskip{Panaiot G. Zotev$^{a*}$, Sam A. Randerson$^a$, Xuerong Hu$^a$, Yue Wang$^b$, Alexander I. Tartakovskii$^{a**}$}
\vskip0.5\baselineskip\footnotesize{\em$^a$Department of Physics and Astronomy, University of Sheffield, Sheffield, S3 7RH, UK\\$^b$School of Physics, Engineering and Technology, University of York, York, YO10 5DD, UK\\}
$^{*}$p.zotev@sheffield.ac.uk \quad $^{**}$a.tartakovskii@sheffield.ac.uk

\end{center}
\vskip1.0\baselineskip

\begin{multicols}{2}

\textbf{Transition metal dichalcogenide (TMD) single photon emitters (SPEs) offer numerous advantages to quantum information applications, such as high single photon purity and deterministic positioning. Strain in the host monolayer, induced by underlying dielectric Mie resonators, is known to localize their formation to positions co-located with near-field photonic hotspots providing further control over their optical properties. However, traditional materials used for the fabrication of nanoresonators, such as silicon or gallium phosphide (GaP), often require a high refractive index substrate resulting in losses of the emitted light and limited photonic enhancement. Here, we use nanoantennas (NAs) fabricated from multilayer TMDs, which allow complete flexibility with the choice of substrate due to the adhesive van der Waals forces, enabling high refractive index contrast or the use of highly reflective metallic surfaces. We demonstrate the localized formation of SPEs in WSe$_2$ monolayers transferred onto WS$_2$ NAs on both SiO$_2$ and Au substrates, enabling strong photonic enhancements and increased single photon collection. We provide evidence for enhanced quantum efficiencies (QE) reaching an average value of 43\% (7\%) for SPEs on WS$_2$ NAs on a SiO$_2$ (Au) substrate. We further combine the advantages offered by both dielectric and metallic substrates to numerically simulate an optimized NA geometry for maximum WSe$_2$ single photon excitation, emission, collection. Thus, the fluorescence is enhanced by a factor of over 4 orders of magnitude compared to vacuum and 5 orders of magnitude compared to a flat SiO$_2$/Si surface. Our work showcases the advantages offered by employing TMD material nanoresonators on various substrates for SPE formation and photonic enhancement.}\\

TMD material single photon emitters have drawn much interest due to their expected use in atomically thin quantum devices which can be deterministically fabricated and integrated with arbitrary substrates, offering narrow linewidths and high single photon purity \cite{Turunen2022}. The formation mechanism of these quantum emitters, discovered first in monolayer WSe$_2$ \cite{Tonndorf2015,He2015,Srivastava2015,Koperski2015,Chakraborty2015} and often associated with strain \cite{Kumar2015,Branny2017a,Rosenberger2019,Daveau2020,Yucel2024}, is still debated. However, this has lead to the proposal of several theoretical models describing their occurrence in cryogenic photoluminescence (PL) including strain-induced potential traps for excitons \cite{Feierabend2019}, various defects \cite{Linhart2019,Dass2019}, both of which are believed to be filled by trapped dark excitons \cite{Branny2017a,Feierabend2019,Linhart2019,Sortino2021} or via long-lived intermediate states \cite{Oreszczuk2020}, as well as momentum-dark states\cite{Lindlau2018}. Despite the lack of definitive knowledge concerning their origin, many steps have been taken to integrate TMD SPEs with devices by employing focused ion beam defect formation \cite{Klein2020} or, more commonly, strain-engineering \cite{Kern2016,Shepard2017,Rosenberger2019,Daveau2020,Yucel2024} and thereby coupling to waveguides \cite{Blauth2018,Peyskens2019}, microcavities \cite{Flatten2018,Drawer2023}, circular Bragg gratings \cite{Iff2021a} and Mie resonators \cite{Luo2018,Cai2018a,Sortino2021}. Additionally, tuning of the properties of the SPEs has been achieved by shaping the strain profile to orient the dipole of the emitter \cite{Kern2016,Luo2020,Paralikis2024}, increasing strain values \cite{Iff2019,Sortino2021} or using gating \cite{Ripin2023} to tune the emission energy and employing van der Waals ferromagnets to achieve chiral single photon emission \cite{Fawzi2018a}.

Applications in quantum information requiring the scalable and controllable realization of on-chip quantum emitters have led to arrays of 2D SPEs realized by transfer of monolayer WSe$_2$ onto pre-fabricated arrays of nanopillars patterned in photoresist \cite{Branny2017a}, SiO$_2$ \cite{Palacios-Berraquero2017}, GaP \cite{Sortino2021} and Au/Si with a spacer layer of Al$_2$O$_3$ on top \cite{Cai2018a}. The latter two of these materials were employed to enhance SPE intensities by coupling with Mie resonances as a result of the close proximity of maximum strain values and photonic mode hotspots \cite{Cai2018a,Sortino2021}. However, each of these devices required the fabrication of nanopillars or Mie resonators from the substrate material thus achieving little to no refractive index contrast, known to enable strong localization of resonances and thus provide large fluorescence enhancements \cite{Robinson2005}, between nanoresonator and underlying surface. 

Recently, works employing van der Waals (vdW) materials as high-index dielectrics for the fabrication of nanophotonic structures \cite{Verre2019,Froch2019,Zhang2020,Munkhbat2022,Zotev2022,Weber2022} have enabled the realization of nanoresonators on arbitrary substrates achieving a large refractive index contrast and thus tightly confined resonances \cite{Robinson2005,Zotev2023} as well as hybrid dielectric-plasmonic resonances with ultra-high Q factors \cite{Zotev2023,Randerson2024}. These vdW material nanophotonic structures thus enable many applications such as waveguiding \cite{Munkhbat2022,Nonahal2023}, strong light-matter coupling \cite{Verre2019,Weber2022,Zotev2023}, lasing \cite{Sung2022} and Purcell enhancement of emission \cite{Zotev2022,Froch2022,Sortino2023}.

Here, we pattern WS$_2$ multilayer crystals deposited on SiO$_2$ and Au substrates into hexagonal dimer nanoantennas. Through dark field spectroscopy we confirmed the presence of Mie resonances which we employed in the fluorescence enhancement of single photon emitters forming in monolayer WSe$_2$ transferred onto the fabricated nanoresonators. Simulations of the resulting NAs on both substrates yielded strong electric field confinement into photonic hotspots at the top surface of the NAs where the strain in the transferred monolayer is expected to be highest, thus inducing the formation of SPEs. These hotspots are expected to enhance the absorption in the monolayer, increase the QE of single photon emission and improve collection efficiencies by redirecting the emitted light into our optical setup \cite{Sortino2021}. The total maximum fluorescence enhancement factor for a quantum emitter forming within these hotspots compared with one in vacuum, calculated from these three contributions, was simulated to reach as high as 914 (10670) for the SiO$_2$ (Au) substrate. When we compare an SPE within the hotspots of the NA on a SiO$_2$ (Au) substrate to an emitter on a flat 300 nm SiO$_2$/Si wafer surface we calculated a fluorescence enhancement as high as 2012 (23493).

In cryogenic PL measurements, we observed the formation of spatially localized, narrow-linewidth emitters which are shown to emit single photons. We recorded a factor of 2.7 higher SPE intensity for the SiO$_2$ substrate. Using pulsed laser excitation and simulations of the collection efficiency near each NA, we extract a quantum efficiency of each emitter. We define this as the average number of photons emitted per excitation pulse. The SPEs forming on WS$_2$ NAs on a SiO$_2$ (Au) substrate yielded an average QE of 43\% (7\%), which is higher than observed for quantum emitters in an unenhanced or weakly enhanced regime ($\approx$ 1.5-4\% \cite{Luo2018,Sortino2021}). We attribute the very high average QE for the SiO$_2$ substrate to the strong mode confinement achievable with a large refractive index contrast ($\approx$ 2.5). The much lower QE of the emitters forming on NAs fabricated on the Au substrate is attributed to non-radiative decay channels from the SPE state, such as charge transfer to the metallic surface \cite{Zhang2019a}. We confirm this via a reduced average SPE lifetime and the reduction of Auger processes observed in the time resolved PL of monolayer WSe$_2$ on a flat Au substrate.

To surmount the disadvantage of increased non-radiative decay and make use of the larger collection efficiency of a metallic substrate, we further design optimized WS$_2$ NAs on a gold substrate with an insulating spacer of SiO$_2$ or hBN. The insulating layer is used to suppress charge transfer processes while also yielding maximum fluorescence enhancement factors of more than 23700 (33200), compared to vacuum, and more than 157900 (221300), compared to a flat 300 nm SiO$_2$/Si surface, for a WSe$_2$ SPE forming within the hotspots at the outer top vertices of the NAs fabricated on a Au substrate with a SiO$_2$ (hBN) spacer under low power excitation. Our work showcases the advantages offered by heterointegration of TMD SPEs and nanoresonators with different substrates, thereby enabling optimized coupling of enhanced quantum emitters with virtually any material system and device. \\


\large
\textbf{Results}\\
\normalsize

\begin{figure}[ht!]
	\centering
  \includegraphics[width=\linewidth]{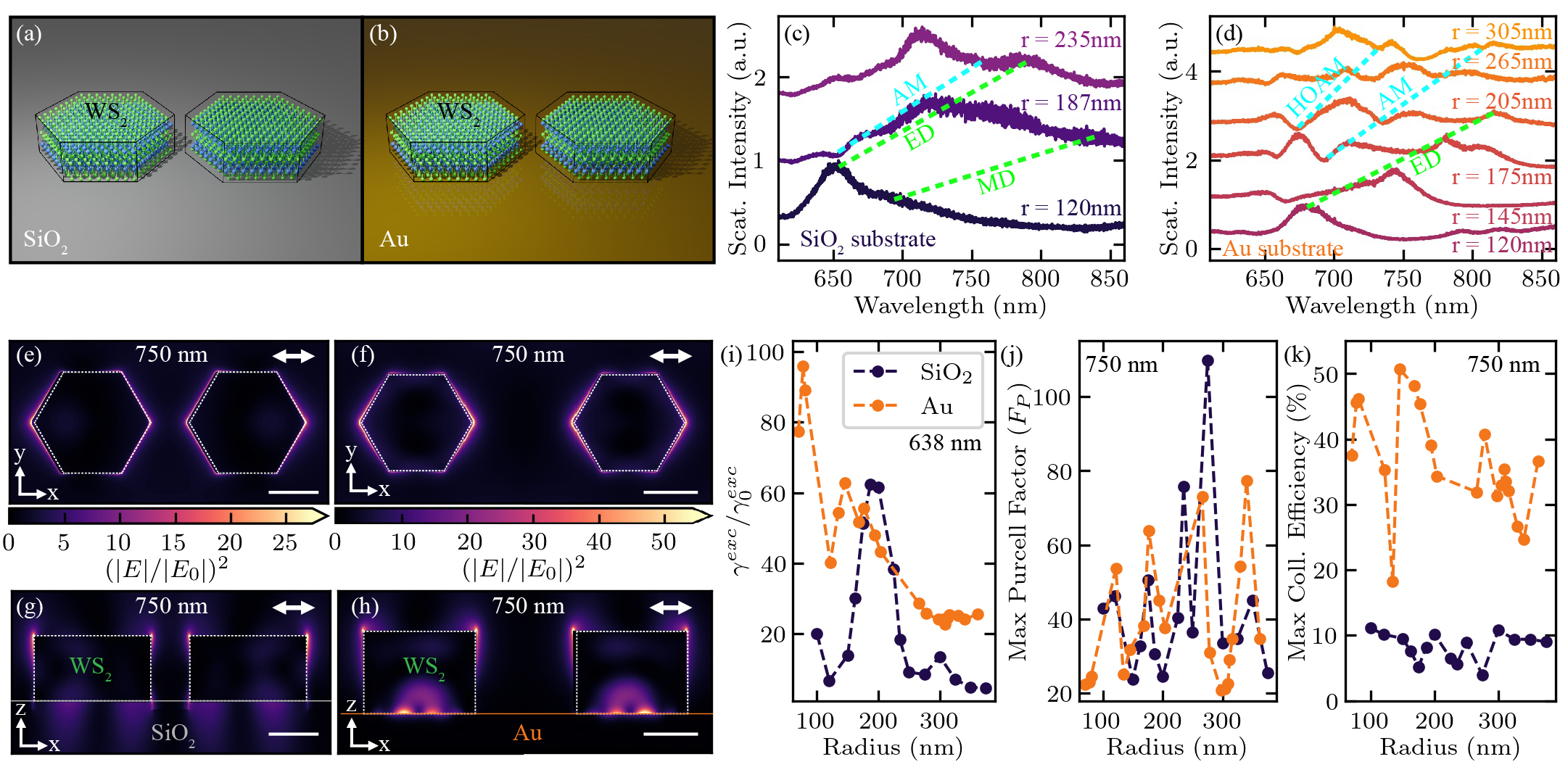}
  \caption{\textbf{Simulation of photonic enhancement expected for WSe$_2$ SPEs forming on WS$_2$ nanoantennas on a SiO$_2$ and Au substrate.}
	\textbf{(a)} and \textbf{(b)} Schematic illustration of WS$_2$ dimer nanoantennas fabricated onto a SiO$_2$ (left) and Au (right) substrate respectively.
	\textbf{(c)} and \textbf{(d)} Experimentally recorded dark field spectra of several WS$_2$ dimer nanoantennas on the SiO$_2$ and Au substrate respectively demonstrating the presence of magnetic (MD) and electric (ED) dipole Mie resonances as well as an anapole (AM) and higher order anapole mode (HOAM) which can be used for photonic enhancement of SPEs.
	\textbf{(e)} and \textbf{(f)} Simulations of the electric field intensity profile at the top surface of a WS$_2$ NA on a SiO$_2$ (r = 235nm, h = 135nm, g = 150nm) and Au (r = 205nm, h = 190nm, g = 515nm) substrate respectively, where WSe$_2$ monolayer SPEs are expected to form after transfer due to the high strain values at the edges of the structure. Electric field hotspots are confined to the the inner and outer vertices of the NA. The wavelength of 750 nm is used as this is near the expected energies of WSe$_2$ SPEs. Scale bars = 200 nm.
	\textbf{(g)} and \textbf{(h)} Simulations of the electric field intensity profile for a vertical cut through the WS$_2$ NA on a SiO$_2$ and Au substrate respectively for the same geometries and wavelength. Electric field hotspots are confined to the top surface of the NA for both substrates, while there is an additional hotspot at interface between the WS$_2$ NA and the Au substrate. 
	\textbf{(i)}, \textbf{(j)} and \textbf{(k)} Excitation rate enhancement factor, Purcell factor, and collection efficiency calculated at appropriate wavelengths for a fabricated array of WS$_2$ NAs with a range of radii on the SiO$_2$ (dark blue) and Au (orange) substrate. }
  \label{F1}
 \end{figure}

\textbf{WS$_2$ nanoantennas on SiO$_2$ and Au substrates} To achieve an entirely TMD SPE system on different substrates, we began by preparing two nominally 290 nm SiO$_2$/Si substrates, one of which was coated with a 100 nm thick Au film sputtered by electron beam. We mechanically exfoliated multilayer WS$_2$ crystals onto these two substrates and employed electron beam lithography and reactive ion etching to pattern hexagonal dimer (double pillar) NAs into the WS$_2$ \cite{Zotev2022,Danielsen2021}. While the SiO$_2$ substrate was slightly overetched (<15 nm), the Au film acted as a natural etch stop \cite{Zotev2023,Randerson2024} leaving 135 nm and 190 nm tall WS$_2$ dimer NAs on the SiO$_2$ and Au substrates respectively (Figure \ref{F1}(a) and (b)). Once fabricated, we characterized the optical resonances of the NAs on both substrates via dark field spectroscopy which yielded the presence of both magnetic and electric dipole Mie modes as well as anapole and higher order anapole resonances (Figures \ref{F1}(c) and (d) for the SiO$_2$ and Au substrate respectively). The height, radius and dimer separation gap for each NA was measured using atomic force microscopy (AFM) and scanning electron microscopy. 

Using the measured geometry of the fabricated WS$_2$ NAs, we simulated the electric field intensity enhancement values surrounding the nanostructures (Figure \ref{F1}(e) and (f) for the SiO$_2$ and Au substrates respectively). The radii of the simulated NAs were 235 nm and 205 nm, their heights were 135 nm and 190 nm and their dimer gaps (distance between individual nanopillars) were 150 nm and 515 nm for the SiO$_2$ and Au substrates respectively. As we aimed to enhance the fluorescence rate of monolayer WSe$_2$ SPEs, emitting near 750 nm, we simulated the expected photonic enhancement at this wavelength and observed confinement of the electric field at the top surface of the NAs. The photonic hotspots at the upper NA vertices are expected to yield higher fluorescence enhancements for the Au substrate. We attribute this to the redistribution of the electric field in the presence of a reflective surface and a reduced leakage into the substrate. As the photonic hotspots at the top surface of the NA are co-located with the highest expected strain values in a monolayer of WSe$_2$ transferred onto these nanostructures \cite{Sortino2020}, we expected a high probability of SPE formation within these hotspots and thus strong enhancement of their fluorescence. In order to calculate a fluorescence enhancement rate compared to vacuum ($F/F_0$), we utilized the following definition \cite{Sortino2019}:

\begin{equation}
\frac{F}{F_0} = \frac{\gamma^{exc}}{\gamma^{exc}_0}\cdot\frac{q^{em}}{q^{em}_0}\cdot\frac{\eta^{em}}{\eta^{em}_0},
\label{eq:1}
\end{equation}

\noindent
where the first factor ($\gamma^{exc}/\gamma^{exc}_0)$ corresponds to the excitation rate enhancement, which is governed by the electric field intensity at the position of the SPE, at the pump laser wavelength. The second factor ($q^{em}/q^{em}_0$) represents the quantum efficiency enhancement at the emission wavelength, which can be approximated by the Purcell factor for emitters with a low intrinsic QE as was demonstrated for WSe$_2$ SPEs in an unenhanced or weakly enhanced regime ($\approx$ 1.5-4\% \cite{Luo2018,Sortino2021}). The last factor ($\eta^{em}/\eta^{em}_0$) corresponds to the light collection efficiency enhancement provided by the photonic structure at the emission wavelength calculated for a numerical aperture of 0.64. In our simulations, each of the enhancement factors is calculated as a comparison to an emitter in a vacuum environment.

To understand the maximum possible fluorescence enhancement for the fabricated NAs on the SiO$_2$ and Au substrates, we simulated the electric field intensity enhancement, Purcell factor and collection efficiency at the appropriate wavelength (638 nm for our laser excitation and 750 nm for the SPE emission wavelength) for a range of radii measured from our fabricated NAs (Figures \ref{F1}(i)-(k)). From the electric field intensity simulations, we observed that for most radii achieved in our fabrication, we can expect stronger excitation rate enhancements for the Au substrate, attributed to the stronger confinement provided by the metallic film, and thus use lower power laser pumping. The Purcell factor simulations yielded similar emission enhancement rates for the two substrates which suggests similar expected QE enhancements in the SPEs on the two substrates. Perhaps the largest advantage of using an Au substrate as opposed to SiO$_2$ is the factor of $\approx$ 4 enhancement of the expected collection efficiency which is also higher than previously achieved on GaP NAs \cite{Sortino2019,Sortino2021}. This is expected from a metallic mirror substrate as much of the light emitted from the SPE toward the substrate can now be reflected towards the collection optics. The maximum fluorescence enhancement factor ($F/F_0$) expected for properly positioned SPEs within the photonic hotspots when compared to vacuum, calculated by substituting in the results from Figures \ref{F1}(i)-(k) into equation \ref{eq:1}, varies from 62.5 to 914 for the SiO$_2$ substrate and from 647.9 to 10670 for the Au substrate over the range of radii. Upon comparison with an emitter forming on a flat 300 nm SiO$_2$/Si surface, the fluorescence enhancements vary from 137 to 2012 for the WS$_2$ NAs on the SiO$_2$ substrate and from 1426 to 23493 for the Au substrate over the same range of radii. \\


\textbf{Single photon emitter formation and quantum efficiency enhancement} To confirm that single photon emitters can be formed by depositing a monolayer TMD onto multilayer TMD nanoresonators fabricated on different substrates, we transfer a single layer of WSe$_2$ onto the fabricated arrays of WS$_2$ NAs on SiO$_2$ and Au substrates (see Supplementary Note 1) via an all-dry polymer stamp technique (see Methods). WS$_2$ NAs on SiO$_2$ have already been shown to enhance the photoluminescence of monolayer WSe$_2$ at room temperature via the Purcell effect \cite{Zotev2022}, thus providing a promising starting point for enhancement of SPEs. 

\begin{figure}[ht!]
	\centering
  \includegraphics[width=\linewidth]{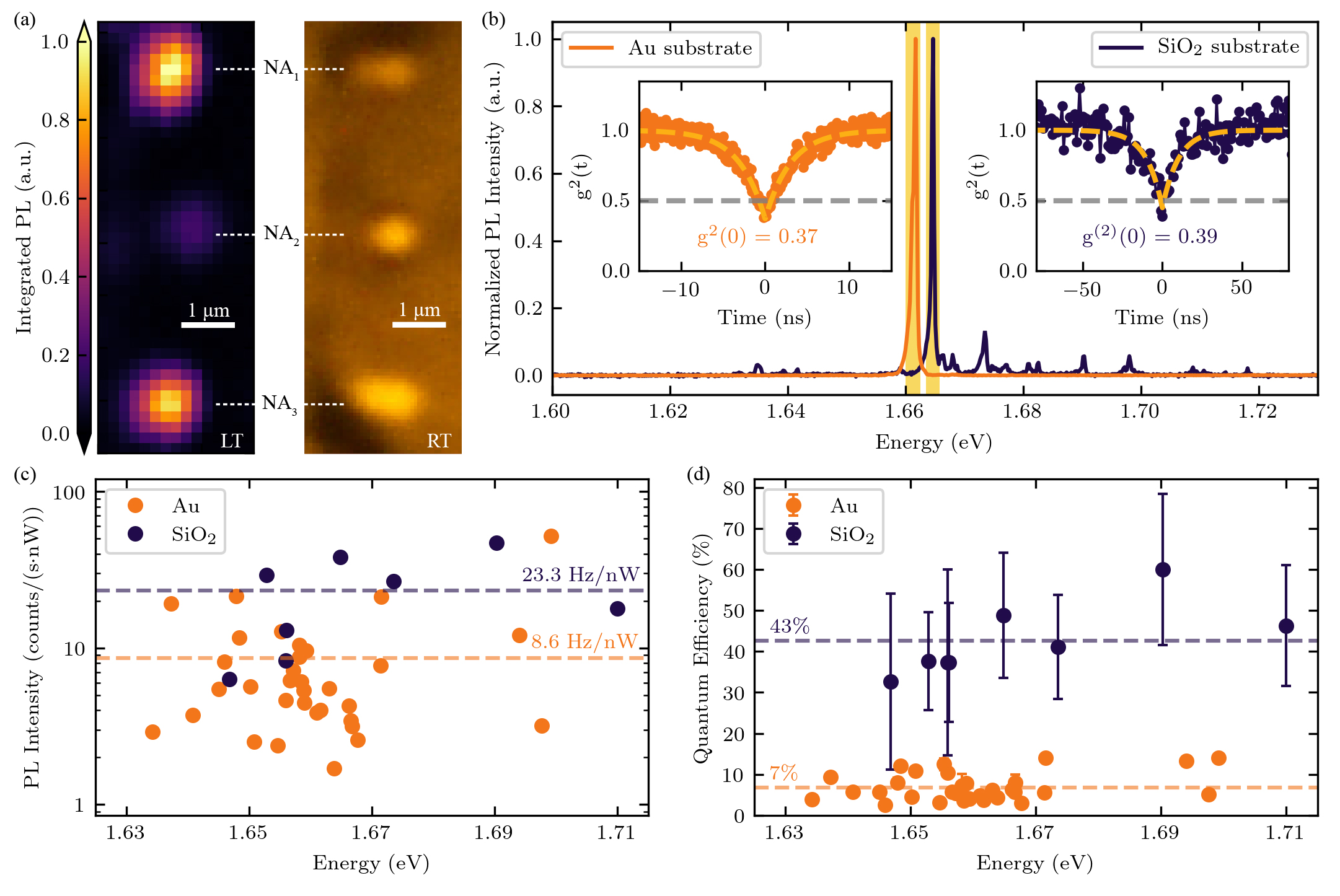}
  \caption{\textbf{WSe$_2$ single photon emitter formation, intensity and quantum efficiency on WS$_2$ NAs on SiO$_2$ and Au substrates.}
	\textbf{(a)} Left Panel: Cryogenic PL map of monolayer WSe$_2$ surrounding three WS$_2$ nanoantenna positions (NA$_1$, NA$_2$, NA$_3$) on the SiO$_2$ substrate integrated from 1.663 to 1.666 eV. Right Panel: Room temperature PL image from the same region of the sample. Overlap of bright emission at both temperatures indicates SPE localization at nanoantenna sites.
	\textbf{(b)} PL spectra of a WSe$_2$ monolayer SPE forming on a WS$_2$ NA fabricated on a SiO$_2$ (dark blue) and Au (orange) substrates exhibiting narrow linewidth emission. Yellow regions highlight integration energies for map in \textbf{(a)} as well as for Hanbury-Brown-Twiss (HBT) experiments. Inset: Results of HBT experiments for both SiO$_2$ (right) and Au (left) substrates exhibiting $g^2(0)$ values below 0.5 indicating single photon emission.
	\textbf{(c)} Integrated intensities per nW of excitation power of SPEs forming on WS$_2$ NAs on SiO$_2$ (dark blue) and Au (orange) substrates at saturation. The SiO$_2$ (Au) substrate sample yielded emitters with an average intensity of 23.3 (8.6) Hz/nW indicated by the horizontal dashed lines.
	\textbf{(d)} Quantum efficiency calculated for each recorded SPE forming on WS$_2$ NAs fabricated on SiO$_2$ and Au substrates. Emitters in SiO$_2$ (Au) substrate sample yielded an average QE of 43\% (7\%) indicated by the horizontal dashed lines. Error bars indicate the uncertainty of the position of the emitter relative to the NA and thus an uncertainty in the collection efficiency used for the QE calculation. Note that the error bars are not visible for some of the SPEs on WS$_2$ NAs on the Au substrate due to a very small difference in the collection efficiency for different emitter positions and the much smaller average QE.}
  \label{F2}
 \end{figure}

Next, we cooled our samples to liquid helium temperature in a bath cryostat and carried out detailed micro-PL experiments. We excited the WSe$_2$ monolayer with a 638 nm pulsed laser (5MHz) and measured PL spectra from different locations at and surrounding the nanoantenna positions. We recorded a spatial map of the emitted PL, integrating between 1.663 and 1.666 eV, from the SiO$_2$ substrate sample near three NAs (Figure \ref{F2}(a), left) and compared it to a room temperature PL image, integrated between 1.24 and 2.25 eV, of the same region (Figure \ref{F2}(a), right). The overlap of the low and room temperature PL in the two panels confirms that the emission is strongly localized to the NA positions. The cryogenic PL spectrum (Figure \ref{F2}(b), blue) recorded from the uppermost NA in the spatial map exhibits an emitter with a linewidth of <500 $\mu$eV, similar to previous reports for QE enhanced SPEs on GaP NAs \cite{Sortino2021}. A similarly narrow emission line was also observed from the WSe$_2$ monolayer transferred onto WS$_2$ NAs on the Au substrate (Figure \ref{F2}(b), orange). We performed Hanbury-Brown-Twiss experiments (Figure \ref{F2}(b), insets) to measure the photon statistics of the recorded lines from the samples with both substrates, resulting in anti-bunching dips below g$^2$(0) = 0.5 at zero time delay, thus confirming the single photon nature of the emission. As expected for 0-D quantum emitters, we also observed linear polarization, power saturation at relatively weak excitation powers (<500 nW) and long lifetimes (7-170 ns) compared to free excitons in the WSe$_2$ monolayer (<4 ps) \cite{Wang2014a,Zhu2014} (see Supplementary Note 2).

We recorded similar narrow-linewidth (see Supplementary Note 2) emission from nearly all of the studied NA positions (>98\%) on both substrates. This provides evidence that the strain induced in the WSe$_2$ monolayer by the WS$_2$ NAs with heights of 135 nm and 190 nm for the SiO$_2$ and Au substrates respectively is sufficient to induce the formation of at least one SPE at each NA position. Interestingly, many of the emitters observed on the Au substrate sample and several of those on the SiO$_2$ surface sample exhibited a polarization along the line connecting the centers of the nanopillars forming the WS$_2$ dimer NA (dimer axis). We attribute this to the strain profile induced in the monolayer during the transfer (see Supplementary Note 2) \cite{Kern2016,Sortino2020}. The higher percentage of emitters forming with a polarization along the dimer axis on Au substrate sample is attributed to the larger strain induced in the monolayer by the taller NAs. This demonstrates an ability to control the polarization of the SPEs and orient it along the NA resonance expected to result in the highest photonic enhancement \cite{Zotev2022} with a more effortless procedure than in previous reports \cite{Luo2020}.

We subsequently recorded the integrated (over the linewidth of the emitter) intensities of many SPEs at saturation forming onto different WS$_2$ NAs on the SiO$_2$ and Au substrates (Figure \ref{F2}(c)). Surprisingly, the average intensity of the emitters on the Au substrate ($\approx$ 8.6 Hz/nW) is 2.7 times lower than that on the SiO$_2$ substrate (23.3 Hz/nW) despite the overall higher excitation rate and collection efficiency enhancements we expected from the simulations in Figures \ref{F1}(i) and (k). We attribute this to the introduction of a very fast non-radiative decay channel, such as charge transfer to the metallic substrate \cite{Zhang2019a}. Room temperature PL and time-resolved PL measurements exhibiting reduced intensities and suppressed exciton-exciton annihilation respectively suggest such a process (see Supplementary Note 3). The reduced average lifetime of emitters recorded for the Au substrate sample as compared to those on the SiO$_2$ surface sample (see Supplementary Note 2) provides further evidence of an increased non-radiative decay rate. We also considered that heating due to the Au substrate absorption may reduce the SPE intensities, however, the local temperature is expected to rise only negligibly (see Supplementary Note 4).

When comparing the intensities of emitters forming on WS$_2$ NAs fabricated on the SiO$_2$ substrate with previously studied WSe$_2$ SPEs forming on SiO$_2$ nanopillars \cite{Palacios-Berraquero2017,Branny2017a,Sortino2021} we observed larger intensities and decay lifetimes leading us to consider enhanced quantum efficiencies. As we utilized pulsed excitation with a lower laser repetition rate than the decay rates of all SPEs, expected to yield an emission rate equal to the repetition rate for a quantum efficiency of 100\%, we can estimate the QE of each emitter via the following:

\begin{equation}
QE = \frac{I}{R\cdot\eta_{em}\cdot\eta_{exp}},
\label{eq:2}
\end{equation}

\noindent
where $I$ is the recorded integrated intensity of the SPE in Hz within the saturation regime, $\eta_{em}$ is the simulated collection efficiency of the objective lens largely depending on the positioning of the emitter relative to the WS$_2$ dimer NA, $\eta_{exp}$ is the measured efficiency of our experimental setup and $R$ is the repetition rate of the excitation laser in Hz. We have measured the efficiency of our experimental setup to be 0.56\% using a 725 nm laser (see Supplementary Note 5). For each emitter, we define a maximum and minimum collection efficiency based on its position relative to the NA which is unknown to us. The largest collection efficiency is found for an SPE position at the upper vertices of the NA and the lowest value is calculated for an emitter in close proximity to the substrate. Utilizing these two values provides us with not only an average estimate of the quantum efficiency of each SPE but also an upper and lower bound (Figure \ref{F2}(d)) depending on the precise position of the emitter relative to the WS$_2$ NA and substrate.

Comparing our current estimate of the QE with previously measured WSe$_2$ SPEs forming on SiO$_2$ nanopillars (4\% \cite{Sortino2021}) and uncoupled gold nanocubes (1.5\% \cite{Luo2018}), we observed an order of magnitude higher quantum efficiencies for emitters forming on WS$_2$ NAs on the SiO$_2$ substrate with an average value of 43\%, which is even higher than observed for quantum efficiency enhanced SPEs forming on coupled gold nanocubes (12.6\% \cite{Luo2018}) and GaP dimer nanoantennas (average $\approx$ 21\% \cite{Sortino2021}). It is worth noting that some quantum efficiencies observed for SPEs forming on GaP nanoantennas reached higher values (86\%), yet the lack of refractive index contrast between nanoantenna and substrate is expected to lead to weaker mode confinement and thus lower average QE enhancement than for the WS$_2$ NAs on the SiO$_2$ substrate. While the emitters forming on WS$_2$ NAs on the Au substrate yielded a much lower average QE of 7\%, this is still higher than that observed on SiO$_2$ nanopillars or uncoupled gold nanocubes, despite the introduction of fast non-radiative charge transfer processes. We attribute this enhanced QE to two factors as discussed in a previous report \cite{Sortino2021}. First, the possibility of low power excitation leads to suppression of non-radiative exciton-exciton annihilation processes \cite{Sortino2021,Mouri2014a}, thereby increasing the QE of the SPEs. This factor is a result of the enhanced absorption at the pump laser wavelength as well as the reduced non-radiative processes (suggested by the long decay times compared to previous reports \cite{He2015,Srivastava2015}) due to either the high crystal quality of the WS$_2$ NAs or SPE formation along suspended portions of the monolayer between contact with the dimer and the substrate. Second, Purcell enhancement of the radiative rate is expected for emitters within the NA hotspots \cite{Zotev2022} also contributing to an increased quantum efficiency.\\

\begin{figure}[ht!]
	\centering
  \includegraphics[width=\linewidth]{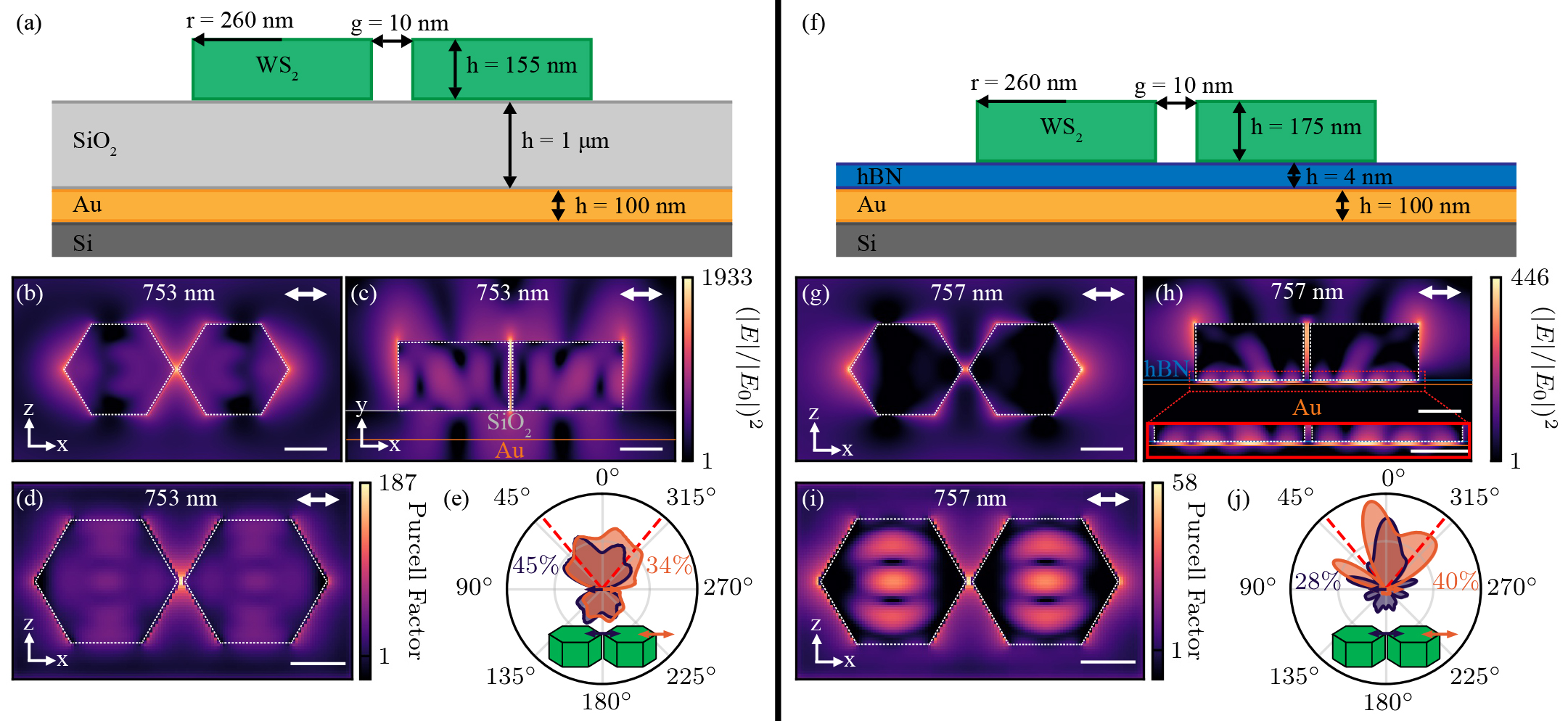}
  \caption{\textbf{Photonic enhancement simulations of optimized WS$_2$ NA designs on a Au substrate with a SiO$_2$ or hBN spacer.}
	\textbf{(a)} and \textbf{(f)} Schematic illustrations of WS$_2$ NAs on a Au substrate with a SiO$_2$ and hBN spacer for optimum fluorescence enhancement of WSe$_2$ monolayer SPEs.
	\textbf{(b)},\textbf{(c)} and \textbf{(g)},\textbf{(h)} Simulated electric field intensity profiles at the top surface and as a vertical cut through the middle of the WS$_2$ NAs for optimized designs with a SiO$_2$ and hBN spacer at the photonic resonance wavelength of 753 nm and 757 nm respectively. Electric field hotspots are observed in the dimer gap and at the outer top vertices of the NAs.
	\textbf{(d)} and \textbf{(i)} Maps of the Purcell factor at different positions at the top surface of the WS$_2$ NA for the SiO$_2$ and hBN spacer designs respectively. Larger enhancements are observed within the dimer gap and at the outer top vertices of the NA, where SPEs are expected to form, as expected from the electric field hotspots in \textbf{(b)},\textbf{(c)},\textbf{(g)} and \textbf{(h)}.
	\textbf{(e)} and \textbf{(j)} Radiation patterns for SPE positions within the dimer gap (dark blue) and at the outer top vertices (orange) for the SiO$_2$ and hBN spacer designs respectively. Emission is directed mainly towards collection optics with the cone defined by the experimental numerical aperture in this study indicated by the red dashed lines. The numbers to the sides indicate the collection efficiency, with a numerical aperture of 0.64, for an emitter forming in the dimer gap (dark blue) and at the outer top vertices of the NA. Lower insets indicate the position of the SPE dipole relative to the WS$_2$ NAs. All scale bars = 200 nm.}
  \label{F3}
\end{figure}

\textbf{Optimized WS$_2$ NAs on a Au substrate with a dielectric spacer} As we have demonstrated so far, WS$_2$ NAs not only enable the photonic enhancement of WSe$_2$ single photon emitters but also offer integration with various substrates providing different advantages such as quantum efficiency or collection efficiency enhancement. To combine the advantages of the two substrates we have studied and suppress charge-transfer processes which limited SPE intensities for the Au substrate sample, we design an optimized photonic structure for maximum enhancement of WSe$_2$ monolayer SPEs forming on WS$_2$ dimer NAs. We consider a gold substrate with a spacer of either sputtered SiO$_2$ (Figure \ref{F3}(a)) or few-layer hBN (Figure \ref{F3}(f)) below the WS$_2$ NAs. We used FDTD simulations to calculate optimized geometries for the two different types of spacers (see Methods) leading to a radius of 260 nm, a height of 155 nm (175 nm), an ultra-small gap of 10 nm achievable through AFM repositioning \cite{Zotev2022} and a spacer layer thickness of 1 $\mu$m (4 nm) for the SiO$_2$ (hBN) spacer. These simulations were carried out with the expectation of an SPE polarization along the WS$_2$ NA dimer axis as we observed this orientation to be more probable due to the strain profile in the transferred monolayer \cite{Kern2016,Sortino2020} and lead to higher mode confinement \cite{Zotev2022}. The photonic resonance wavelengths were also constrained to a range coinciding with the range of discovered SPE energies in WSe$_2$. The optimized hBN spacer is similar in thickness to an Al$_2$O$_3$ buffer layer (6 nm) grown atop plasmonic nanopillars, used for SPE formation and enhancement, to reduce quenching of WSe$_2$ monolayer emission \cite{Cai2018a}. 

We first simulated the photonic hotspots near the upper surface of the NAs (Figure \ref{F3}(b)-(c) and (g)-(h) for the SiO$_2$ and hBN spacer design respectively) leading to excitation rate enhancements within the dimer gap of 1933 (177) for the SiO$_2$ (hBN) spacer design. The strong confinement in the few-layer hBN spacer (Figure \ref{F3}(f)) leads to far lower electric field intensity values at the top surface of the NA when compared to the SiO$_2$ spacer design. However, the maximum electric field intensities at the outer nanoantenna vertices, where the strain in a transferred WSe$_2$ monolayer is expected to be the highest \cite{Sortino2020} and thus most likely to form SPEs, reach far more similar values of 545 and 446 for the SiO$_2$ and hBN spacer design respectively. The simulated electric field intensity enhancements are directly responsible for an increase in the excitation and thus the fluorescence rate if resonant excitation is employed. Precise tuning of the NA resonance wavelength can be achieved by varying the size, leading to only small reductions in the photonic enhancement in our simulations.

Next, we simulated the expected Purcell factors, directly leading to an increase in quantum efficiency and thus fluorescence rate, for WSe$_2$ SPEs forming on the top surface of the WS$_2$ NAs in both designs (Figure \ref{F3}(d) and (i) for the SiO$_2$ and hBN spacer design respectively). These are maximized at the inner NA vertices reaching values of 187 (58) for the SiO$_2$ (hBN) spacer design. While the outer vertices of the NAs, where SPEs are expected to form, yielded similar Purcell factors of 22 and 32 for the SiO$_2$ and hBN spacer designs respectively. Due to the reduced charge-transfer expected for these nanoantenna designs, we also predict far lower non-radiative rates leading to higher SPE QE and intensity than observed for the WSe$_2$ SPEs recorded for the Au substrate sample. 

Lastly we simulated the radiation pattern expected from these optimized designs, which enables the calculation of collection efficiency and thus the fluorescence rate for SPEs forming both in the dimer gap and at the more probable outer vertices of the NAs (Figures \ref{F3}(e) and (j) for the for the SiO$_2$ and hBN spacer design respectively). We observed that most of the emitted light is expected to be redirected away from the Au reflective substrate leading to a collection efficiency of 45\% (28\%) within our experimental numerical aperture (0.64) for an SPE forming within the dimer gap of the SiO$_2$ (hBN) spacer design. At the more probable SPE location atop the outer vertex of the NA, we calculated a collection efficiency of 34\% (40\%) for the SiO$_2$ (hBN) spacer design. Comparing these designs to the collection efficiency of WS$_2$ NAs on the SiO$_2$ substrate used in our experimental study, we expect an enhancement ranging from 2.8 to 4.5.

The fluorescence enhancement we expect to achieve for WSe$_2$ SPEs forming on the optimized SiO$_2$ and hBN spacer designs, when compared to vacuum or a flat SiO$_2$/Si surface, using the same experimental setup and resonant excitation can be calculated using equation \ref{eq:1} after two considerations. First, Purcell enhancement of the second factor in equation \ref{eq:1} will saturate at a factor of 66, assuming an intrinsic, unenhanced SPE quantum yield of 1.5\% \cite{Luo2018}, as the QE cannot reach higher than 100\%. Second, the fluorescence rate will not increase if the excitation rate becomes larger than the lifetime of emission \cite{Cambiasso2018}. These two conditions constrain our definition of the fluorescence enhancement to hold only for low intrinsic QE emitters, such as WSe$_2$ SPEs, and for low laser pumping powers resulting in unenhanced excitation rates far below the emission rate. As the quantum efficiency of SPEs forming in some of the photonic hotspots of the optimized designs may reach 100\%, the Purcell enhancement may lead to shorter emission lifetimes and enable even higher laser repetition rates than used in our experiments.

Under these conditions, the fluorescence enhancement factor compared to vacuum ($F/F_0$) for WSe$_2$ SPEs forming within the dimer gap of the optimized SiO$_2$ (hBN) spacer design may reach more than 5 (4) orders of magnitude. Upon comparison with an emitter on a flat 300 nm SiO$_2$/Si surface the fluorescence enhancement factor may reach more than 6 (5) orders of magnitude. The far more probable location for SPE formation atop an outside vertex of the NAs yielded maximum fluorescence enhancements of more than 23700 (33200) compared to vacuum for the SiO$_2$ (hBN) spacer designs which is higher than simulated for the NAs on both substrates studied in experiment. When comparing to an emitter forming on a flat 300 nm SiO$_2$/Si surface the fluorescence enhancement may reach factors of more than 157900 (221300). 

To enable a more simple fabrication of such WS$_2$ NAs on a reflective substrate with an insulating spacer, we also simulated designs with a gap of 50 nm, achievable without AFM repositioning. These yielded smaller electric field intensity enhancements and Purcell factors, however, the collection efficiencies were similar leading to a similar expectation of strong fluorescence from WSe$_2$ SPEs (see Supplementary Note 6). \\


\FloatBarrier

\large
\textbf{Discussion}\\
\normalsize

Our study of WSe$_2$ monolayer SPEs forming on WS$_2$ nanoantennas on different substrates illustrates the advantages offered by employing vdW nanoresonators for the photonic enhancement of quantum emitters. We demonstrated the feasibility of achieving vdW nanophotonic structures on both dielectric and metallic surfaces by fabricating WS$_2$ dimer NAs on SiO$_2$ and Au substrates which were confirmed to host Mie resonances. FDTD simulations of the fabricated NAs yielded pump laser absorption, quantum efficiency and collection efficiency enhancements for single photon emitters formed within their resonant hotspots. 

We transferred monolayers of WSe$_2$ on the WS$_2$ NAs fabricated on both substrates and observed the formation of spatially localized, narrow-linewidth single photon emitters in cryogenic PL measurements which exhibited power saturation and long lifetimes indicative of a 0-D quantum state. The linear polarization of most of the emitters forming on the NAs on the Au substrate was discovered to be oriented along the dimer axis which we attribute to the strain profile induced in the monolayer during transfer \cite{Kern2016,Sortino2020}. This provides a more effortless control of the polarization orientation of the SPE than in previous reports \cite{Luo2020}. By studying the emission intensity at saturation under pulsed excitation, we were able to calculate the QE of each SPE enabling us to demonstrate enhancements for the WS$_2$ NAs on a SiO$_2$ substrate with an average quantum efficiency of 43\%, which is higher than previous reports on GaP NAs \cite{Sortino2021}, Au nanocubes \cite{Luo2018} and more than an order of magnitude higher than on SiO$_2$ nanopillars \cite{Sortino2021}. We attribute this enhancement to the strong resonance confinement enabled by the large refractive index contrast ($\approx$ 2.5) achieved when fabricating WS$_2$ NAs on a SiO$_2$ substrate leading to lower saturation powers and enhanced Purcell factors \cite{Robinson2005} resulting in increased quantum efficiencies. The QE of SPEs forming on WS$_2$ NAs on the Au substrate (7\%) were also higher than those forming on SiO$_2$ nanopillars \cite{Sortino2021} and uncoupled gold nanocubes \cite{Luo2018}, providing evidence of photonic enhancement, despite the introduction of a non-radiative decay mechanism through charge transfer. 

The study of WSe$_2$ SPEs forming on WS$_2$ NAs on different surfaces enabled us to further design and propose optimized nanoantenna geometries on Au substrates with an hBN or SiO$_2$ spacer layer. Such structures employ the advantages of strong light confinement due to large refractive index contrast \cite{Robinson2005} as well as improved collection efficiencies due to the reflective substrate, therefore enabling maximum photonic enhancement and collection of the single photon emission. These improved geometries were simulated to yield electric field intensity enhancements up to 1933, Purcell factors of more than 180 and collection efficiencies as high as 45\%, leading to maximum expected fluorescence enhancements of more than 6 orders of magnitude for WSe$_2$ SPEs forming within the dimer gap of the hexagonal WS$_2$ dimer NA, under low pumping conditions. For far more likely positioned SPEs on the outer vertices of the NAs, we calculated a similarly strong maximum fluorescence enhancement of more than 5 orders of magnitude. The strong electric field intensity and Purcell enhancements are also expected to enable the use of lower pumping powers with higher repetition rates than used in our experiments in order to excite WSe$_2$ SPEs.

Our study illustrates the advantages offered to TMD SPE formation and enhancement by vdW material nanoresonators which can be fabricated on virtually any substrate further enabling a blend of approaches for maximized photonic enhancement. This material platform enables simple and scalable integration of bright TMD SPEs with devices used in a variety of applications including quantum cryptography and information processing. \\


\large
\textbf{Methods}\\ 
\normalsize

\textbf{Sample fabrication}

\textit{Gold substrate preparation}: In order to prepare the gold substrate, we firstly deposit a 10 nm layer of Ti onto a 290nm SiO$_2$/Si substrate via e-beam evaporation in order to improve the adhesion between substrate and gold. We subsequently deposit 100 nm of gold via the same method.

\textit{Van der Waals materials exfoliation}: WS$_2$ crystals are mechanically exfoliated from bulk (HQ-graphene) onto a nominally 290 nm SiO$_{2}$ on silicon or Au substrate. Large crystals with recognizable axes via straight edged sides at 120$^{\circ}$ to each other were identified and their positions within the sample were recorded for further patterning.

\textit{Electron beam lithography}: Samples were spin coated with ARP-9 resist (AllResist GmbH) at 3500 rpm for 60 s and baked at 180$^\circ$ for 5 min yielding a film of 200 nm thickness. Electron beam lithography was performed in a Raith GmbH Voyager system operating at 50 kV using a beam current of 560 pA.

\textit{Reactive ion etching}: The recipe used for dry etching included SF$_{6}$ (30 sccm) at a DC bias of 40 V and a pressure of 0.13 mbar for 40 seconds. Removal of the remaining resist after etching was accomplished by a bath in warm 1165 resist remover (1 hour) followed by Acetone (5 min) and IPA (5 min). If resist is still found on the sample, final cleaning is done in a bath of Acetone (1 hour) and IPA (5 min) followed by 1 hour in a UV ozone treatment. In some cases, the structures were slightly over-etched leading to NAs with a small pedestal of SiO$_2$ (<15 nm) or gold (<5 nm). This, however, did not lead to any noticeable changes in the photonic resonances.

\textit{WSe$_2$ transfer}: WSe$_{2}$ monolayers were mechanically exfoliated from a bulk crystal (HQ-graphene) onto a (PDMS) stamp, which had previously been attached to a glass slide. Large monolayers were identified using PL imaging. The glass slide is rotated upside down and attached to a holder arm by means of a vacuum. The target substrate, consisting of WS$_2$ nanoantennas on a SiO$_2$ or Au surface, was also held to a stage using the same vacuum. The WSe$_2$ monolayer was slowly brought into contact with the target substrate through the use of a piezo-scanner stage. After the entire monolayer has contacted the surface, the glass slide with PDMS was slowly moved away from the target substrate. The low speed of the peeling process makes use of the visco-elastic properties of the PDMS polymer and leaves the monolayer of WSe$_2$ onto the substrate. \\

\textbf{Dark field spectroscopy}
Optical spectroscopy in a dark-field configuration was achieved using a Nikon LV150N microscope with a fiber-coupled output. Incident illumination from a tungsten halogen lamp in the microscope was guided to a circular beam block with a diameter smaller than the beam diameter. The light was then reflected by a 45$^\circ$ tilted annular mirror towards a 50x Nikon (0.8 NA) dark-field objective which only illuminates the sample at large angles to the normal. Reflected light from the sample is guided back through the same objective towards a fiber coupler. Due to the small diameter of the multimode fiber core used, only light reflected back at small angles to the normal is collected. The fiber from the microscope was subsequently coupled to a Princeton Instruments spectrometer and charge coupled device. \\

\textbf{Micro-Photoluminescence spectroscopy}
In order to record the photoluminescence emitted from monolayer WSe$_2$ at different regions of our sample, we used a home-built setup, which includes a pulsed diode laser at 638 nm. The sample was mounted into an attocube bath cryostat insert following which the chamber was pumped to vacuum and inserted into a helium bath. Heat exchange was achieved by a small amount of helium gas being let into the insert. The collimated excitation laser was passed through a 700 nm short-pass filter, a Glan-Thompson linear polarizer and a half wave plate before being deflected by a 50:50 beam-splitter and passing through the insert window followed by an aspheric lens (0.64 numerical aperture) which focused the beam onto the sample. The emitted light is collected by the same lens and passes through the insert window, beam-splitter a half wave plate, thin film polarizer and a 700 nm long-pass filter to be guided through a single mode fiber before being focused onto the slit of a Princeton Instruments spectrometer (0.75 meter) and CCD. The PL lifetime studies utilized the pulsed laser excitation and the emitted light was spectrally filtered (10 nm) using the exit slit of the spectrometer before it was fiber-coupled to an ID Quantique avalanche photo-diode (id100). \\

\textbf{Hanbury-Brown-Twiss experiment}
The autocorrelation measurements used to provide evidence of single photon emission were performed using the same setup as for the micro-photoluminescence studies. However, the 700 nm long-pass filter was exchanged for a 750 nm bandpass filter (spectral width = 10 nm) and the fiber output was coupled to a superconducting single photon detector (Single Quantum). The relatively high values of g$^2$(0) are largely attributed to the poor bandpass filter utilized to measure the photon statistics of the SPEs. \\

\textbf{FDTD simulations}
The finite-difference time-domain simulations were carried out using Lumerical Inc. software. The geometry of hexagonal WS$_2$ nanoantennas on a SiO$_2$ or Au substrate with and without a spacer were defined within the software utilizing the refractive index of WS$_2$ from reference \cite{Zotev2023}, SiO$_2$ from reference \cite{Palik1998}, Au from reference \cite{Johnson1972} and hBN from reference \cite{Zotev2023}.

\textit{Electric field intensity simulations}: Calculations of the electric field intensity normalized to vacuum were simulated using plane wave illumination propagating normal to the surface using a TFSF source from the air side. The illumination was polarized along the dimer axis with a plane monitor recording the electric field 0.5 nm above the top surface of the nanoantennas or as a vertical cross-section of the structure passing through the dimer axis.

\textit{Purcell factor simulations}: Simulations of the Purcell factor normalized to vacuum were carried out using the same geometry as for the electric field intensity simulations. The illumination was achieved through a dipole source placed at different positions, 0.5 nm above the top surface of the nanoantenna with a polarization parallel to the dimer axis. 

\textit{Collection efficiency simulations}: Calculations of the collection efficiency were carried out using the same geometry as before. Illumination was achieved via a dipole source placed 0.5 nm above either an inside vertex of the WS$_2$ NA, an outside vertex of the NA, or the substrate including SiO$_2$ and Au. The power of the emitted light was collected using a infinite plane monitor placed 0.5 nm above the dipole source following which a filter corresponding to the 0.64 numerical aperture was applied to the collected power. This power was subsequently normalized to the collected power in all directions from the dipole. Collection efficiency enhancement was then calculated by normalizing the collection efficiency for a dipole within the NA geometry to one in vacuum. \\

\textit{Optimization procedure for WS$_2$ NAs on a Au substrate with a dielectric spacer}: In order to optimize the geometry of the WS$_2$ NAs on the Au substrate with a SiO$_2$ or hbN spacer, simulations of the electric field intensity were performed, as confinement of the electric field is expected to lead to Purcell enhancement. As the electric field intensity and Purcell factor have far larger contributions to the fluorescence enhancement, which was the main focus of this procedure, the collection efficiency was only calculated after electric field optimization. The electric field intensity was recorded for the inner and outer vertices of the WS$_2$ NAs where SPE formation is probable. The radius and height of the nanoantenna as well as the thickness of the spacer layer were varied consecutively until a maximum electric field intensity was achieved within the photonic hotspots. Simulations for the two spacer designs with a gap of 10 nm and 50 nm were all carried out individually. As the initial optimization procedure yielded several spacer thicknesses resulting in similar electric field intensities, the collection efficiency for each was used to determine the optimum thickness. \\


\large
\textbf{Acknowledgments}\\
\normalsize


\large
\textbf{Author Contributions}\\
\normalsize

X.H. and Y.W. deposited the metal to form the Au substrate. X.H. and P.G.Z. exfoliated WS$_2$ onto the SiO$_2$ and Au substrates. X.H. and Y.W. both contributed to the fabrication of the WS$_2$ nanoantennas on both substrates. P.G.Z. and S.R. carried out the scanning electron microscopy and AFM measurements to characterize the geometry of the fabricated nanoantennas. P.G.Z. and S.R. transferred the WS$_2$ monolayer onto the fabricated NAs on both substrates and performed the dark-field and micro-photoluminescence measurements. P.G.Z. carried out the Hanbury-Brown-Twiss experiments. P.G.Z. and A.I.T. conceived the experiments and simulations as well as analyzed the results and wrote the manuscript with contributions from all co-authors. A.I.T. oversaw the entire project. \\


\bibliographystyle{unsrt}
\bibliography{./library}

\begin{thebibliography}{10}

\bibitem{Turunen2022}
Mikko Turunen, Mauro Brotons-Gisbert, Yunyun Dai, Yadong Wang, Eleanor Scerri,
  Cristian Bonato, Klaus~D. Jöns, Zhipei Sun, and Brian~D. Gerardot.
\newblock Quantum photonics with layered 2d materials.
\newblock {\em Nature Reviews Physics}, 0123456789, 2022.

\bibitem{Tonndorf2015}
Philipp Tonndorf, Robert Schmidt, Robert Schneider, Johannes Kern, Michele
  Buscema, Gary~A. Steele, Andres Castellanos-Gomez, Herre S.~J. van~der Zant,
  Steffen~Michaelis de~Vasconcellos, and Rudolf Bratschitsch.
\newblock Single-photon emission from localized excitons in an atomically thin
  semiconductor.
\newblock {\em Optica}, 2:347, 2015.

\bibitem{He2015}
Yu-Ming He, Genevieve Clark, John~R. Schaibley, Yu-Ming He, Ming-Cheng Chen,
  Yu-Jia Wei, Xing Ding, Qiang Zhang, Wang Yao, Xiaodong Xu, Chao-Yang Lu, and
  Jian-Wei Pan.
\newblock Single quantum emitters in monolayer semiconductors.
\newblock {\em Nature Nanotechnology}, 10:497--502, 2015.

\bibitem{Srivastava2015}
Ajit Srivastava, Meinrad Sidler, Adrien~V. Allain, Dominik~S. Lembke, Andras
  Kis, and A.~Imamoğlu.
\newblock Optically active quantum dots in monolayer wse$_2$.
\newblock {\em Nature Nanotechnology}, 10:491--496, 2015.

\bibitem{Koperski2015}
M.~Koperski, K.~Nogajewski, A.~Arora, J.~Marcus, P.~Kossacki, and M.~Potemski.
\newblock Single photon emitters in exfoliated wse$_2$ structures.
\newblock {\em Nature Nanotechnology}, 10:503--506, 2015.

\bibitem{Chakraborty2015}
Chitraleema Chakraborty, Laura Kinnischtzke, Kenneth~M. Goodfellow, Ryan Beams,
  and A.~Nick Vamivakas.
\newblock Voltage-controlled quantum light from an atomically thin
  semiconductor.
\newblock {\em Nature Nanotechnology}, 10:507--511, 2015.

\bibitem{Kumar2015}
S.~Kumar, A.~Kaczmarczyk, and B.~D. Gerardot.
\newblock Strain-induced spatial and spectral isolation of quantum emitters in
  mono- and bilayer wse$_2$.
\newblock {\em Nano Letters}, 15:7567--7573, 2015.

\bibitem{Branny2017a}
Artur Branny, Santosh Kumar, Raphaël Proux, and Brian~D. Gerardot.
\newblock Deterministic strain-induced arrays of quantum emitters in a
  two-dimensional semiconductor.
\newblock {\em Nature Communications}, 8:15053, 2017.

\bibitem{Rosenberger2019}
Matthew~R. Rosenberger, Chandriker~Kavir Dass, Hsun~Jen Chuang, Saujan~V.
  Sivaram, Kathleen~M. McCreary, Joshua~R. Hendrickson, and Berend~T. Jonker.
\newblock Quantum calligraphy: Writing single-photon emitters in a
  two-dimensional materials platform.
\newblock {\em ACS Nano}, 13:904--912, 2019.

\bibitem{Daveau2020}
Raphaël~S. Daveau, Tom Vandekerckhove, Arunabh Mukherjee, Zefang Wang, Jie
  Shan, Kin~Fai Mak, A.~Nick Vamivakas, Gregory~D. Fuchs, A.~Nick Vamivakas,
  and Gregory~D. Fuchs.
\newblock Spectral and spatial isolation of single tungsten diselenide quantum
  emitters using hexagonal boron nitride wrinkles.
\newblock {\em APL Photonics}, 5:096105, 2020.

\bibitem{Yucel2024}
Oguzhan Yücel, Denis Yagodkin, Jan~N. Kirchhof, Abhijeet Kumar, Adrian
  Dewambrechies, Sviatoslav Kovalchuk, Yufeng Yu, and Kirill~I. Bolotin.
\newblock Strain activation of localized states in wse$_2$.
\newblock {\em arXiv}, 2 2024.

\bibitem{Feierabend2019}
Maja Feierabend, Samuel Brem, and Ermin Malic.
\newblock Optical fingerprint of bright and dark localized excitonic states in
  atomically thin 2d materials.
\newblock {\em Physical Chemistry Chemical Physics}, 21:26077--26083, 2019.

\bibitem{Linhart2019}
Lukas Linhart, Matthias Paur, Valerie Smejkal, Joachim Burgdörfer, Thomas
  Mueller, and Florian Libisch.
\newblock Localized inter-valley defect excitons as single-photon emitters in
  wse$_2$.
\newblock {\em Physical Review Letters}, 123:146401, 2019.

\bibitem{Dass2019}
Chandriker~Kavir Dass, Mahtab~A. Khan, Genevieve Clark, Jeffrey~A. Simon, Ricky
  Gibson, Shin Mou, Xiaodong Xu, Michael~N. Leuenberger, and Joshua~R.
  Hendrickson.
\newblock Ultra‐long lifetimes of single quantum emitters in monolayer
  wse$_2$/hbn heterostructures.
\newblock {\em Advanced Quantum Technologies}, 1900022:1900022, 2019.

\bibitem{Sortino2021}
Luca Sortino, Panaiot~G. Zotev, Catherine~L. Phillips, Alistair~J. Brash,
  Javier Cambiasso, Elena Marensi, A.~Mark Fox, Stefan~A. Maier, Riccardo
  Sapienza, and Alexander~I. Tartakovskii.
\newblock Bright single photon emitters with enhanced quantum efficiency in a
  two-dimensional semiconductor coupled with dielectric nano-antennas.
\newblock {\em Nature Communications}, 12:6063, 2021.

\bibitem{Oreszczuk2020}
K.~Oreszczuk, T.~Kazimierczuk, T.~Smoleński, K.~Nogajewski, M.~Grzeszczyk,
  A.~Łopion, M.~Potemski, and P.~Kossacki.
\newblock Carrier relaxation to quantum emitters in few-layer wse$_2$.
\newblock {\em Physical Review B}, 102:1--5, 2020.

\bibitem{Lindlau2018}
Jessica Lindlau, Malte Selig, Andre Neumann, Léo Colombier, Jonathan Förste,
  Victor Funk, Michael Förg, Jonghwan Kim, Gunnar Berghäuser, Takashi
  Taniguchi, Kenji Watanabe, Feng Wang, Ermin Malic, and Alexander Högele.
\newblock The role of momentum-dark excitons in the elementary optical response
  of bilayer wse$_2$.
\newblock {\em Nature Communications}, 9:1--7, 2018.

\bibitem{Klein2020}
Authors~J Klein, L~Sigl, S~Gyger, K~Barthelmi, M~Florian, S~Rey, and
  T~Taniguchi.
\newblock Scalable single-photon sources in atomically thin mos$_2$.
\newblock {\em arXiv}, 2020.

\bibitem{Kern2016}
Johannes Kern, Iris Niehues, Philipp Tonndorf, Robert Schmidt, Daniel Wigger,
  Robert Schneider, Torsten Stiehm, Steffen~Michaelis de~Vasconcellos, Doris~E.
  Reiter, Tilmann Kuhn, and Rudolf Bratschitsch.
\newblock Nanoscale positioning of single-photon emitters in atomically thin
  wse$_2$.
\newblock {\em Advanced Materials}, 28:7101--7105, 2016.

\bibitem{Shepard2017}
Gabriella~D. Shepard, Obafunso~A. Ajayi, Xiangzhi Li, X.~Y. Zhu, James Hone,
  and Stefan Strauf.
\newblock Nanobubble induced formation of quantum emitters in monolayer
  semiconductors.
\newblock {\em 2D Materials}, 4:021019, 2017.

\bibitem{Blauth2018}
Mäx Blauth, Marius Jürgensen, Gwenaëlle Vest, Oliver Hartwig, Maximilian
  Prechtl, John Cerne, Jonathan~J. Finley, and Michael Kaniber.
\newblock Coupling single photons from discrete quantum emitters in wse$_2$ to
  lithographically defined plasmonic slot-waveguides.
\newblock {\em Nano Letters}, 18:6812–6819, 2018.

\bibitem{Peyskens2019}
Frédéric Peyskens, Chitraleema Chakraborty, Muhammad Muneeb, Dries~Van
  Thourhout, and Dirk Englund.
\newblock Integration of single photon emitters in 2d layered materials with a
  silicon nitride photonic chip.
\newblock {\em Nature Communications}, 10, 12 2019.

\bibitem{Flatten2018}
L.~C. Flatten, L.~Weng, A.~Branny, S.~Johnson, P.~R. Dolan, A.~A.P.~P Trichet,
  B.~D. Gerardot, and J.~M. Smith.
\newblock Microcavity enhanced single photon emission from two-dimensional
  wse$_2$.
\newblock {\em Applied Physics Letters}, 112, 2018.

\bibitem{Drawer2023}
J~C Drawer, V~N Mitryakhin, H~Shan, S~Stephan, M~Gittinger, L~Lackner, and
  B~Han.
\newblock Ultra-bright single photon source based on an atomically thin
  material.
\newblock {\em arXiv}, 2023.

\bibitem{Iff2021a}
Oliver Iff, Quirin Buchinger, Magdalena Moczała-Dusanowska, Martin Kamp, Simon
  Betzold, Marcelo Davanco, Kartik Srinivasan, Sefaattin Tongay, Carlos
  Antón-Solanas, Sven Höfling, and Christian Schneider.
\newblock Purcell-enhanced single photon source based on a deterministically
  placed wse2monolayer quantum dot in a circular bragg grating cavity.
\newblock {\em Nano Letters}, 21:4715--4720, 2021.

\bibitem{Luo2018}
Yue Luo, Gabriella~D. Shepard, Jenny~V. Ardelean, Daniel~A. Rhodes, Bumho Kim,
  Katayun Barmak, James~C. Hone, and Stefan Strauf.
\newblock Deterministic coupling of site-controlled quantum emitters in
  monolayer wse$_2$ to plasmonic nanocavities.
\newblock {\em Nature Nanotechnology}, 13:1137–1142, 2018.

\bibitem{Cai2018a}
Tao Cai, Je~Hyung Kim, Zhili Yang, Subhojit Dutta, Shahriar Aghaeimeibodi, and
  Edo Waks.
\newblock Radiative enhancement of single quantum emitters in wse$_2$
  monolayers using site-controlled metallic nanopillars.
\newblock {\em ACS Photonics}, 5:3466--3471, 2018.

\bibitem{Luo2020}
Yue Luo, Na~Liu, Bumho Kim, James Hone, and Stefan Strauf.
\newblock Exciton dipole orientation of strain-induced quantum emitters in
  wse$_2$.
\newblock {\em Nano Letters}, 20:5119--5126, 2020.

\bibitem{Paralikis2024}
Athanasios Paralikis, Claudia Piccinini, Abdulmalik~A. Madigawa, Pietro Metuh,
  Luca Vannucci, Niels Gregersen, and Battulga Munkhbat.
\newblock Tailoring polarization in wse$_2$ quantum emitters through
  deterministic strain engineering.
\newblock {\em arXiv}, 2 2024.

\bibitem{Iff2019}
Oliver Iff, Davide Tedeschi, Javier Martín-Sánchez, Magdalena
  Moczała-Dusanowska, Sefaattin Tongay, Kentaro Yumigeta, Javier
  Taboada-Gutiérrez, Matteo Savaresi, Armando Rastelli, Pablo
  Alonso-González, Sven Höfling, Rinaldo Trotta, and Christian Schneider.
\newblock Strain-tunable single photon sources in wse$_2$ monolayers.
\newblock {\em Nano Letters}, 19:6931--6936, 2019.

\bibitem{Ripin2023}
Adina Ripin, Ruoming Peng, Xiaowei Zhang, Srivatsa Chakravarthi, and Minhao He.
\newblock Tunable phononic coupling in excitonic quantum emitters.
\newblock {\em arXiv}, pages 1--20, 2023.

\bibitem{Fawzi2018a}
Na~Liu, Licheng Xiao, Shichen Fu, Yichen Ma, Song Liu, Siwei Chen, James Hone,
  Eui-Hyeok Yang, and Stefan Strauf.
\newblock Chiral single photons from deterministic quantum emitter arrays via
  proximity coupling to van der waals ferromagnets.
\newblock {\em 2D Materials}, pages 0--23, 2018.

\bibitem{Palacios-Berraquero2017}
Carmen Palacios-Berraquero, Dhiren~M. Kara, Alejandro R.~P. Montblanch, Matteo
  Barbone, Pawel Latawiec, Duhee Yoon, Anna~K. Ott, Marko Loncar, Andrea~C.
  Ferrari, and Mete Atature.
\newblock Large-scale quantum-emitter arrays in atomically thin semiconductors.
\newblock {\em Nature Communications}, 8:15093, 2017.

\bibitem{Robinson2005}
Jacob~T. Robinson, Christina Manolatou, Long Chen, and Michal Lipson.
\newblock Ultrasmall mode volumes in dielectric optical microcavities.
\newblock {\em Physical Review Letters}, 95:1--4, 2005.

\bibitem{Verre2019}
Ruggero Verre, Denis~G. Baranov, Battulga Munkhbat, Jorge Cuadra, Mikael Käll,
  and Timur Shegai.
\newblock Transition metal dichalcogenide nanodisks as high-index dielectric
  mie nanoresonators.
\newblock {\em Nature Nanotechnology}, 14:679--683, 2019.

\bibitem{Froch2019}
Johannes~E. Fröch, Yongsop Hwang, Sejeong Kim, Igor Aharonovich, and Milos
  Toth.
\newblock Photonic nanostructures from hexagonal boron nitride.
\newblock {\em Advanced Optical Materials}, 7:1801344, 2019.

\bibitem{Zhang2020}
Xingwang Zhang, Xiaojie Zhang, Wenzhuo Huang, Kedi Wu, Mengqiang Zhao,
  A.~T.~Charlie Johnson, Sefaattin Tongay, and Ertugrul Cubukcu.
\newblock Ultrathin ws$_2$-on-glass photonic crystal for self-resonant
  exciton-polaritonics.
\newblock {\em Advanced Optical Materials}, 8:1901988, 2020.

\bibitem{Munkhbat2022}
Battulga Munkhbat, Betül Küçüköz, Denis~G. Baranov, Tomasz~J. Antosiewicz,
  and Timur~O. Shegai.
\newblock Nanostructured transition metal dichalcogenide multilayers for
  advanced nanophotonics.
\newblock {\em Laser and Photonics Reviews}, page 2200057, 2022.

\bibitem{Zotev2022}
Panaiot~G Zotev, Yue Wang, Luca Sortino, Toby~Severs Millard, Nic Mullin,
  Donato Conteduca, Mostafa Shagar, Armando Genco, Jamie~K Hobbs, Thomas~F
  Krauss, and Alexander~I Tartakovskii.
\newblock Transition metal dichalcogenide dimer nanoantennas for tailored
  light-matter interactions.
\newblock {\em ACS Nano}, 16:6493--6505, 2022.

\bibitem{Weber2022}
Thomas Weber, Lucca Kühner, Luca Sortino, Amine~Ben Mhenni, Nathan~P. Wilson,
  Julius Kühne, Jonathan~J. Finley, Stefan~A. Maier, and Andreas Tittl.
\newblock Intrinsic strong light-matter coupling with self-hybridized bound
  states in the continuum in van der waals metasurfaces.
\newblock {\em Nature Materials}, 22:970--976, 2023.

\bibitem{Zotev2023}
Panaiot~G. Zotev, Yue Wang, Daniel Andres-penares, Toby~Severs Millard, Sam
  Randerson, Xuerong Hu, Luca Sortino, Charalambos Louca, Mauro
  Brotons-Gisbert, Tahiyat Huq, Stefano Vezzoli, Riccardo Sapienza, Thomas~F.
  Krauss, Brian~D Gerardot, Alexander~I. Tartakovskii, Stefano Vezzoli,
  Riccardo Sapienza, Thomas~F. Krauss, Brian~D Gerardot, and Alexander~I.
  Tartakovskii.
\newblock Van der waals materials for applications in nanophotonics.
\newblock {\em Laser and Photonics Reviews}, 17:2200957, 2023.

\bibitem{Randerson2024}
Sam~A. Randerson, Panaiot~G. Zotev, Xuerong Hu, Alexander~J. Knight, Yadong
  Wang, Sharada Nagarkar, Dominic Hensman, Yue Wang, and Alexander~I.
  Tartakovskii.
\newblock High q hybrid mie–plasmonic resonances in van der waals
  nanoantennas on gold substrate.
\newblock {\em ACS Nano}, 6 2024.

\bibitem{Nonahal2023}
Milad Nonahal, Chi Li, Haoran Ren, Lesley Spencer, Mehran Kianinia, Milos Toth,
  and Igor Aharonovich.
\newblock Engineering quantum nanophotonic components from hexagonal boron
  nitride.
\newblock {\em Laser and Photonics Reviews}, 17:2300019, 2023.

\bibitem{Sung2022}
Junghyun Sung, Dongjin Shin, Hyunhee~Hee Cho, Seong~Won Lee, Seungmin Park,
  Young~Duck Kim, Jong~Sung Moon, Je~hyung Hyung~Kim, and Su~hyun Hyun~Gong.
\newblock Room-temperature continuous-wave indirect-bandgap transition lasing
  in an ultra-thin ws2 disk.
\newblock {\em Nature Photonics}, 16:792--797, 2022.

\bibitem{Froch2022}
Johannes~E. Fröch, Chi Li, Yongliang Chen, Milos Toth, Mehran Kianinia,
  Sejeong Kim, and Igor Aharonovich.
\newblock Purcell enhancement of a cavity-coupled emitter in hexagonal boron
  nitride.
\newblock {\em Small}, 18:2104805, 2022.

\bibitem{Sortino2023}
Luca Sortino, Angus Gale, Lucca Kühner, Chi Li, Jonas Biechteler, Fedja~J.
  Wendisch, Mehran Kianinia, Haoran Ren, Milos Toth, Stefan~A. Maier, Igor
  Aharonovich, and Andreas Tittl.
\newblock Optically addressable spin defects coupled to bound states in the
  continuum metasurfaces.
\newblock {\em Nature Communications}, 15:2008, 2023.

\bibitem{Zhang2019a}
Linglong Zhang, Han Yan, Xueqian Sun, Miheng Dong, Tanju Yildirim, Bowen Wang,
  Bo~Wen, Guru~Prakash Neupane, Ankur Sharma, Yi~Zhu, Jian Zhang, Kun Liang,
  Boqing Liu, Hieu~T. Nguyen, Daniel Macdonald, and Yuerui Lu.
\newblock Modulated interlayer charge transfer dynamics in a monolayer
  tmd/metal junction.
\newblock {\em Nanoscale}, 11:418--425, 2019.

\bibitem{Danielsen2021}
Dorte~R. Danielsen, Anton Lyksborg-Andersen, Kirstine~E.S. Nielsen, Bjarke~S.
  Jessen, Timothy~J. Booth, Manh~Ha Doan, Yingqiu Zhou, Peter Bøggild, and
  Lene Gammelgaard.
\newblock Super-resolution nanolithography of two-dimensional materials by
  anisotropic etching.
\newblock {\em ACS Applied Materials and Interfaces}, 13:41886--41894, 2021.

\bibitem{Sortino2020}
Luca Sortino, Matthew Brooks, Panaiot~G. Zotev, Armando Genco, Javier
  Cambiasso, Sandro Mignuzzi, Stefan~A. Maier, Guido Burkard, Riccardo
  Sapienza, and Alexander~I. Tartakovskii.
\newblock Dielectric nano-antennas for strain engineering in atomically thin
  two-dimensional semiconductors.
\newblock {\em ACS Photonics}, 7:2413--2422, 2020.

\bibitem{Sortino2019}
L.~Sortino, P.~G. Zotev, S.~Mignuzzi, J.~Cambiasso, D.~Schmidt, A.~Genco,
  M.~Aßmann, M.~Bayer, S.~A. Maier, R.~Sapienza, and A.~I. Tartakovskii.
\newblock Enhanced light-matter interaction in an atomically thin semiconductor
  coupled with dielectric nano-antennas.
\newblock {\em Nature Communications}, 10:5119, 2019.

\bibitem{Wang2014a}
G.~Wang, L.~Bouet, D.~Lagarde, M.~Vidal, A.~Balocchi, T.~Amand, X.~Marie, and
  B.~Urbaszek.
\newblock Valley dynamics probed through charged and neutral exciton emission
  in monolayer wse$_2$.
\newblock {\em Physical Review B - Condensed Matter and Materials Physics},
  90:075413, 2014.

\bibitem{Zhu2014}
C~R Zhu, K~Zhang, M~Glazov, B~Urbaszek, T~Amand, Z~W Ji, B~L Liu, and X~Marie.
\newblock Exciton valley dynamics probed by kerr rotation in wse$_2$
  monolayers.
\newblock {\em Physical Review B}, 161302:1--5, 2014.

\bibitem{Mouri2014a}
Shinichiro Mouri, Yuhei Miyauchi, Minglin Toh, Weijie Zhao, Goki Eda, and
  Kazunari Matsuda.
\newblock Nonlinear photoluminescence in atomically thin layered wse$_2$
  arising from diffusion-assisted exciton-exciton annihilation.
\newblock {\em Physical Review B - Condensed Matter and Materials Physics},
  90:155449, 2014.

\bibitem{Cambiasso2018}
Javier Cambiasso, Matthias König, Emiliano Cortés, Sebastian Schlücker, and
  Stefan~A. Maier.
\newblock Surface-enhanced spectroscopies of a molecular monolayer in an
  all-dielectric nanoantenna.
\newblock {\em ACS Photonics}, 5:1546--1557, 2018.

\bibitem{Palik1998}
Edward~D. Palik.
\newblock Handbook of optical constants of solids.
\newblock {\em Academic Press}, I-III, 1998.

\bibitem{Johnson1972}
P~B Johnson and R~W Christy.
\newblock Optical constants of the noble metals.
\newblock {\em Phys. Rev. B}, 6:4370--4379, 1972.

\end{thebibliography}


\begin{thebibliography}{10}

\bibitem{Zotev2022}
Panaiot~G Zotev, Yue Wang, Luca Sortino, Toby~Severs Millard, Nic Mullin,
  Donato Conteduca, Mostafa Shagar, Armando Genco, Jamie~K Hobbs, Thomas~F
  Krauss, and Alexander~I Tartakovskii.
\newblock Transition metal dichalcogenide dimer nanoantennas for tailored
  light-matter interactions.
\newblock {\em ACS Nano}, 16:6493--6505, 2022.

\bibitem{Sortino2020}
Luca Sortino, Matthew Brooks, Panaiot~G. Zotev, Armando Genco, Javier
  Cambiasso, Sandro Mignuzzi, Stefan~A. Maier, Guido Burkard, Riccardo
  Sapienza, and Alexander~I. Tartakovskii.
\newblock Dielectric nano-antennas for strain engineering in atomically thin
  two-dimensional semiconductors.
\newblock {\em ACS Photonics}, 7:2413--2422, 2020.

\bibitem{Zhang2019a}
Linglong Zhang, Han Yan, Xueqian Sun, Miheng Dong, Tanju Yildirim, Bowen Wang,
  Bo~Wen, Guru~Prakash Neupane, Ankur Sharma, Yi~Zhu, Jian Zhang, Kun Liang,
  Boqing Liu, Hieu~T. Nguyen, Daniel Macdonald, and Yuerui Lu.
\newblock Modulated interlayer charge transfer dynamics in a monolayer
  tmd/metal junction.
\newblock {\em Nanoscale}, 11:418--425, 2019.

\bibitem{He2015}
Yu-Ming He, Genevieve Clark, John~R. Schaibley, Yu-Ming He, Ming-Cheng Chen,
  Yu-Jia Wei, Xing Ding, Qiang Zhang, Wang Yao, Xiaodong Xu, Chao-Yang Lu, and
  Jian-Wei Pan.
\newblock Single quantum emitters in monolayer semiconductors.
\newblock {\em Nature Nanotechnology}, 10:497--502, 2015.

\bibitem{Chakraborty2015}
Chitraleema Chakraborty, Laura Kinnischtzke, Kenneth~M. Goodfellow, Ryan Beams,
  and A.~Nick Vamivakas.
\newblock Voltage-controlled quantum light from an atomically thin
  semiconductor.
\newblock {\em Nature Nanotechnology}, 10:507--511, 2015.

\bibitem{Srivastava2015}
Ajit Srivastava, Meinrad Sidler, Adrien~V. Allain, Dominik~S. Lembke, Andras
  Kis, and A.~Imamoğlu.
\newblock Optically active quantum dots in monolayer wse$_2$.
\newblock {\em Nature Nanotechnology}, 10:491--496, 2015.

\bibitem{Kumar2015}
S.~Kumar, A.~Kaczmarczyk, and B.~D. Gerardot.
\newblock Strain-induced spatial and spectral isolation of quantum emitters in
  mono- and bilayer wse$_2$.
\newblock {\em Nano Letters}, 15:7567--7573, 2015.

\bibitem{Dass2019}
Chandriker~Kavir Dass, Mahtab~A. Khan, Genevieve Clark, Jeffrey~A. Simon, Ricky
  Gibson, Shin Mou, Xiaodong Xu, Michael~N. Leuenberger, and Joshua~R.
  Hendrickson.
\newblock Ultra‐long lifetimes of single quantum emitters in monolayer
  wse$_2$/hbn heterostructures.
\newblock {\em Advanced Quantum Technologies}, 1900022:1900022, 2019.

\bibitem{Sortino2021}
Luca Sortino, Panaiot~G. Zotev, Catherine~L. Phillips, Alistair~J. Brash,
  Javier Cambiasso, Elena Marensi, A.~Mark Fox, Stefan~A. Maier, Riccardo
  Sapienza, and Alexander~I. Tartakovskii.
\newblock Bright single photon emitters with enhanced quantum efficiency in a
  two-dimensional semiconductor coupled with dielectric nano-antennas.
\newblock {\em Nature Communications}, 12:6063, 2021.

\bibitem{Mouri2014a}
Shinichiro Mouri, Yuhei Miyauchi, Minglin Toh, Weijie Zhao, Goki Eda, and
  Kazunari Matsuda.
\newblock Nonlinear photoluminescence in atomically thin layered wse$_2$
  arising from diffusion-assisted exciton-exciton annihilation.
\newblock {\em Physical Review B - Condensed Matter and Materials Physics},
  90:155449, 2014.

\bibitem{Cadiz2018}
F.~Cadiz, C.~Robert, E.~Courtade, M.~Manca, L.~Martinelli, T.~Taniguchi,
  T.~Amand, A.~C.H.~H Rowe, D.~Paget, B.~Urbaszek, X.~Marie, K.~Watanabe,
  T.~Amand, A.~C.H.~H Rowe, D.~Paget, B.~Urbaszek, and X.~Marie.
\newblock Exciton diffusion in wse$_2$ monolayers embedded in a van der waals
  heterostructure.
\newblock {\em Applied Physics Letters}, 112:152106, 2018.

\bibitem{Luo2019}
Yue Luo, Na~Liu, Xiangzhi Li, James~C. Hone, and Stefan Strauf.
\newblock Single photon emission in wse2 up 160 k by quantum yield control.
\newblock {\em 2D Materials}, 6, 5 2019.

\bibitem{Johnson1972}
P~B Johnson and R~W Christy.
\newblock Optical constants of the noble metals.
\newblock {\em Phys. Rev. B}, 6:4370--4379, 1972.

\bibitem{Zotev2023}
Panaiot~G. Zotev, Yue Wang, Daniel Andres-penares, Toby~Severs Millard, Sam
  Randerson, Xuerong Hu, Luca Sortino, Charalambos Louca, Mauro
  Brotons-Gisbert, Tahiyat Huq, Stefano Vezzoli, Riccardo Sapienza, Thomas~F.
  Krauss, Brian~D Gerardot, Alexander~I. Tartakovskii, Stefano Vezzoli,
  Riccardo Sapienza, Thomas~F. Krauss, Brian~D Gerardot, and Alexander~I.
  Tartakovskii.
\newblock Van der waals materials for applications in nanophotonics.
\newblock {\em Laser and Photonics Reviews}, 17:2200957, 2023.

\bibitem{Baffou2011}
Guillaume Baffou and Hervé Rigneault.
\newblock Femtosecond-pulsed optical heating of gold nanoparticles.
\newblock {\em Physical Review B - Condensed Matter and Materials Physics},
  84:1--13, 2011.

\bibitem{Baffou2010}
Guillaume Baffou, Romain Quidant, and F.~Javier García~De Abajo.
\newblock Nanoscale control of optical heating in complex plasmonic systems.
\newblock {\em ACS Nano}, 4:709--716, 2010.

\end{thebibliography}

\end{multicols}

\pagebreak

\FloatBarrier

\FloatBarrier

\end{document}


\onehalfspacing
\pagenumbering{arabic}

\begin{center}
{\Large{\bf {\Large{\bf Supplementary Information for: Single photon emitters in monolayer semiconductors coupled to transition metal dichalcogenide nanoantennas on silica and gold substrates}}}}
\vskip1.0\baselineskip{Panaiot G. Zotev$^{a*}$, Sam Randerson$^a$, Xuerong Hu$^a$, Yue Wang$^b$, Alexander I. Tartakovskii$^{a**}$}
\vskip0.5\baselineskip\footnotesize{\em$^a$Department of Physics and Astronomy, University of Sheffield, Sheffield, S3 7RH, UK\\$^b$School of Physics, Engineering and Technology, University of York, York, YO10 5DD, UK\\}
$^{*}$p.zotev@sheffield.ac.uk \quad $^{**}$a.tartakovskii@sheffield.ac.uk

\end{center}
\vskip1.0\baselineskip


\setcounter{figure}{0}
\renewcommand{\thefigure}{S\arabic{figure}}

\begin{center}
\Large
\textbf{Supplementary Note 1: Monolayer transfer onto nanoantennas fabricated on SiO$_2$ and Au substrate}
\end{center}

\begin{figure}[hb!]
	\centering
  \includegraphics[width=\linewidth]{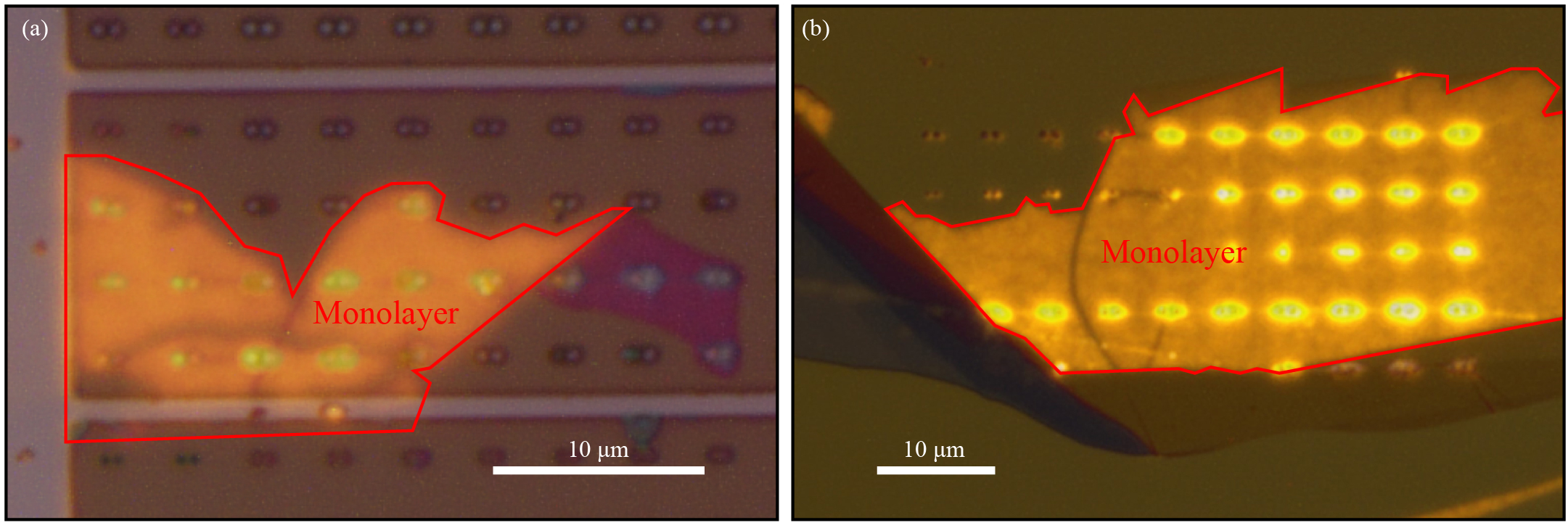}
  \caption{\textbf{Microscope images of WSe$_2$ monolayer transferred onto WS$_2$ NAs on a SiO$_2$ and Au substrate.} 
	\textbf{(a)} and \textbf{(b)} Bright field microscope images of a WSe$_2$ monolayer transferred onto WS$_2$ NAs on a SiO$_2$ and Au substrate respectively. Each image is overlayed with a room temperature PL image taken in the same microscope. Bright positions surrounding the WS$_2$ NAs provide evidence of fluorescence enhancement of the WSe$_2$ monolayer emission. The monolayer region is outlined in red.
	\textbf{(b)}.}
\label{SF1}
\end{figure}

We used an all dry PDMS polymer stamp technique to mechanically transfer exfoliated monolayers of WSe$_2$ onto arrays of WS$_2$ nanoantennas (NAs) which were previously fabricated on SiO$_2$ and Au substrates. Figures \ref{SF1}(a) and (b) display optical microscopy images of the completed samples overlayed by PL images at room temperature. Brightening of the PL surrounding the NA positions is expected as a result of a photonic enhancement due to coupling of the WSe$_2$ emission to the NA resonances \cite{Zotev2022}.

\FloatBarrier
\pagebreak

\begin{center}
\Large
\textbf{Supplementary Note 2: Properties of single photon emitters forming on WS$_2$ nanoantennas on SiO$_2$ and Au substrates}
\end{center}

\begin{figure}[ht!]
	\centering
  \includegraphics[width=\linewidth]{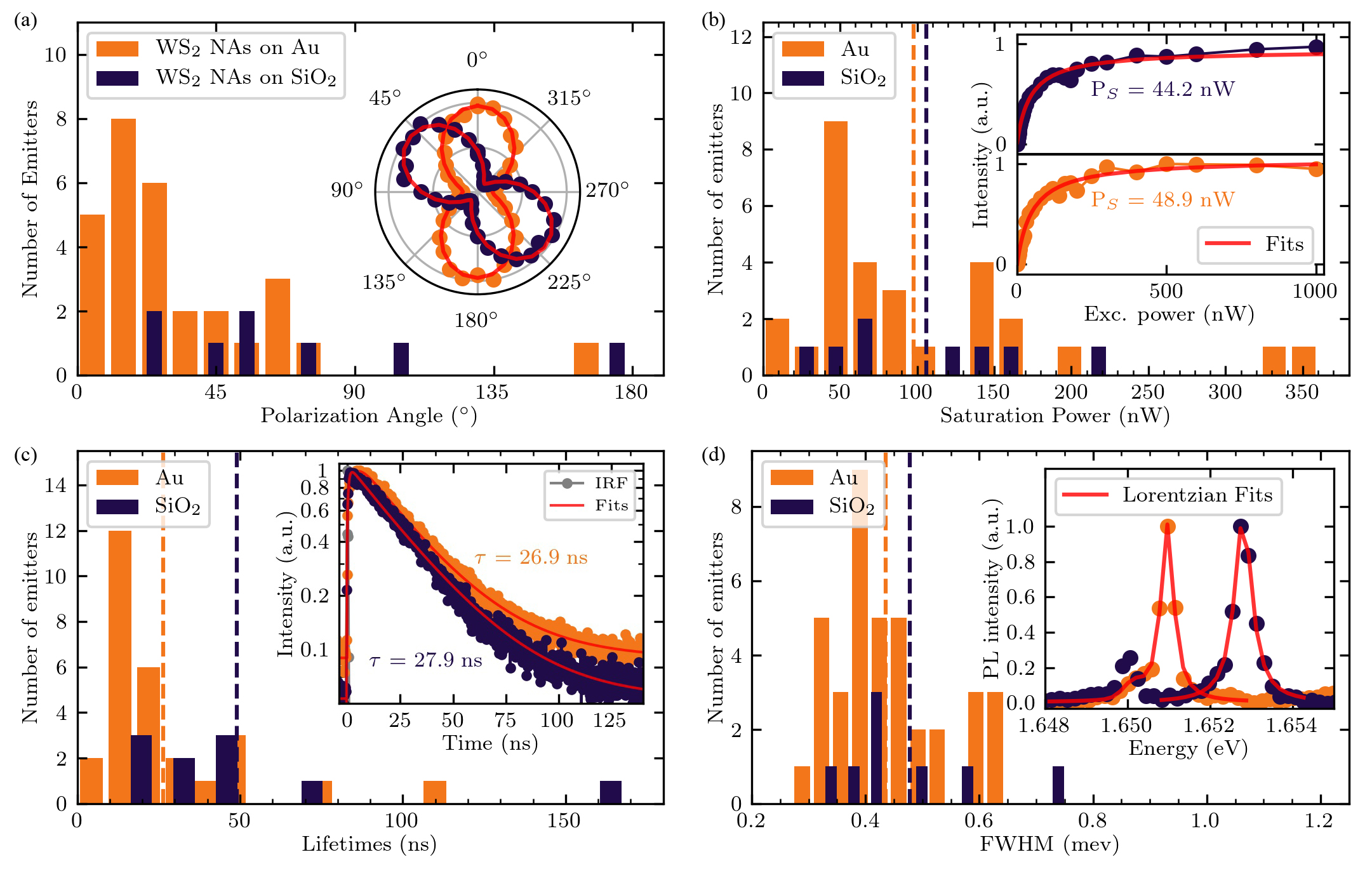}
  \caption{\textbf{Statistics of linear polarization, power saturation, PL decay lifetime and linewidth recorded for SPEs on the SiO$_2$ and Au substrates.} 
	\textbf{(a)} Histogram of linear polarization angles recorded for SPEs on WS$_2$ NAs on the SiO$_2$ (dark blue) and Au (orange) substrates. Many emitters were found to exhibit a polarization parallel or nearly parallel to the dimer axis (0$^{\circ}$. Inset: Exemplary cos$^2$(x) fits of SPE emission intensity with varying polarization angle indicative of linear polarization.
	\textbf{(b)} Histogram of excitation powers at which SPE intensities saturate for emitters on WS$_2$ NAs on the SiO$_2$ (dark blue) and Au (orange) substrates. The average values recorded for the SiO$_2$ (Au) substrate sample are 106.03 nW (97.58 nW) as indicated by the vertical dashed lines. Inset: Exemplary power dependent intensities of emitters on both substrates with similar saturation powers extracted from power law fits.
	\textbf{(c)} Histogram of PL decay lifetimes recorded for SPEs on WS$_2$ NAs on the SiO$_2$ (dark blue) and Au (orange) substrates. The average values recorded for the SiO$_2$ (Au) substrate sample are 49 ns (26.35 ns) as indicated by the vertical dashed lines. The lower lifetimes for the Au substrate sample suggest the presence of an additional non-radiative decay process. Inset: Exemplary SPE PL decay traces fitted with single exponential decay curves from which the emission lifetime is extracted.
	\textbf{(d)} Histogram of linewidths recorded for SPEs on WS$_2$ NAs on the SiO$_2$ (dark blue) and Au (orange) substrates. The average values recorded for the SiO$_2$ (Au) substrate sample are 478$\mu$eV (436$\mu$eV) as indicated by the vertical dashed lines. Inset: Exemplary SPE emission spectra fitted with Lorentzian functions.}
\label{SFigPolPwLifeLineW}
\end{figure}

While studying the photoluminescence of the WSe$_2$ monolayer transferred on WS$_2$ nanoantennas (NAs) fabricated on both SiO$_2$ and Au substrates, we recorded a statistics of several different properties of the formed single photon emitters. Firstly, by varying a half-wave plate in the detection path of our setup, we recorded a clear cos$^2$(x) shape confirming linear polarization of the single photon emitters (SPE) emission as shown in the inset of Figure \ref{SFigPolPwLifeLineW}(a). As we have set the polarization angle of 0$^\circ$ to be parallel to the axis connecting the centers of the nanopillars in the WS$_2$ dimer nanoantenna, we note that a majority of the SPEs recorded for the Au substrate appear to closely conform to this angle. This suggests that many of the emitters form with a dipole moment along this dimer axis which is attributed to the underlying nanoantenna geometry and the strain profile applied to the monolayer during transfer \cite{Sortino2020}. This confirms that control over the polarization angle of the SPEs can be achieved by using appropriately designed nanoresonators.

Secondly, we increased the power of our excitation laser (638 nm, Pulsed: 5Mhz) and observed the saturation of the emission intensity of all of the studied emitters on both substrates as expected for filled 0-D quantum states, as shown in the inset of Figure \ref{SFigPolPwLifeLineW}(b). Each power dependent measurement was fit to an equation of the form $I = I_{sat}\cdot P/(P + P_{sat})$ where $I_{sat}$ indicates the maximum intensity of the SPE and $P_{sat}$ indicates the power at which saturation begins to occur. We recorded very similar powers for the onset of saturation for both SPEs forming on WS$_2$ NAs on the SiO$_2$ and Au substrates with averages of 106.03 nW and 97.58 nW respectively as displayed in Figure \ref{SFigPolPwLifeLineW}(b). We expect that additional decay channels, such as charge transfer to the metallic film \cite{Zhang2019a}, may lead to higher saturation powers for the Au substrate as a result of a reduced exciton trapping for similar excitation powers. However, a suppression of Auger processes due to the reduced exciton population from charge transfer to the substrate would lead to lower saturation powers and thus counteract this effect. 

Due to the additional charge transfer decay channel, we expected that the emission lifetime of the SPEs forming on the NAs on the Au film would exhibit reduced lifetimes when compared to those on the SiO$_2$ substrate as a result of an increase in the non-radiative decay rate. This was observed in the lifetime statistics we collected from emitters forming on NAs on both substrates as displayed in Figure \ref{SFigPolPwLifeLineW}(c). The average lifetime of the studied SPEs on the Au film was recorded to be 26.35 ns which is nearly a factor of 2 smaller than for the SPEs forming on NAs on the SiO$_2$ substrate where the average recorded lifetime was 49 ns. The measured lifetimes, examples of which are plotted in the inset of Figure \ref{SFigPolPwLifeLineW}(c), were also often found to be longer than observed in many previous reports \cite{He2015,Chakraborty2015,Srivastava2015,Kumar2015} reaching values of 100s of ns. A few previous works on SPEs in hBN encapsulated WSe$_2$ monolayers \cite{Dass2019} and SPEs forming on GaP dimer nanoantennas \cite{Sortino2021} also observed such long emission lifetimes which were attributed to a reduced non-radiative rate, and thus higher quantum efficiency, thereby providing nearly-pure measurements of the native WSe$_2$ SPE radiative decay time which may be much longer than previously believed.

Lastly, while studying the PL spectra of the quantum emitters, examples of which are plotted in the inset of Figure \ref{SFigPolPwLifeLineW}(d), we recorded a statistics of their full width at half maxima (FWHM) which are plotted in Figure \ref{SFigPolPwLifeLineW}(d). The average linewidth of the SPEs on the Au film (436$\mu$eV) were found to be slightly lower than for those on the SiO$_2$ substrate (478$\mu$eV). This correlates with an increased non-radiative decay rate observed from PL dynamics measurements as the reduced lifetime is expected to yield less spectral wandering and thus a reduction in the recorded FWHM.

\FloatBarrier

\begin{center}
\Large
\textbf{Supplementary Note 3: Room temperature PL and emission decay from monolayer WSe$_2$ on a SiO$_2$ and Au substrate}
\end{center}

\begin{figure}[hb!]
	\centering
  \includegraphics[width=\linewidth]{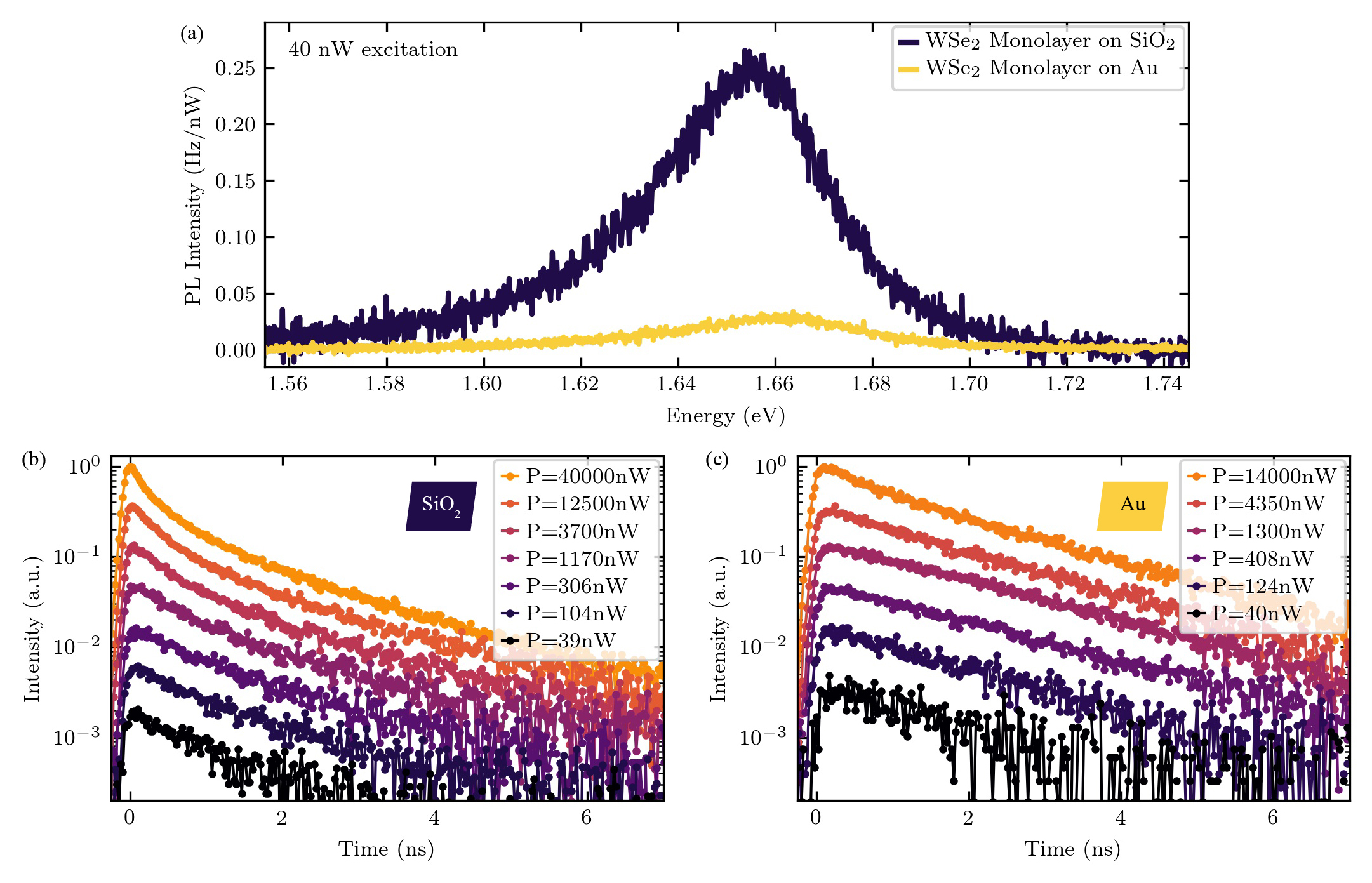}
  \caption{\textbf{Room temperature PL spectra and power dependent emission decay.}
	\textbf{(a)} Room temperature PL spectra recorded for a WSe$_2$ monolayer transferred onto a flat SiO$_2$ (dark blue) and Au (orange) substrate under 40 nW of pulsed (5 MHz) excitation at 638 nm.
	\textbf{(b)} Power dependent PL decay traces for monolayer WSe$_2$ emission on a flat SiO$_2$ substrate. Single exponential decay is observed for excitation powers below 125 nW. Double exponential decay indicative of Auger recombination is present at highter excitation powers.
	\textbf{(c)} Power dependent PL decay traces for monolayer WSe$_2$ emission on a flat Au substrate. Single exponential decay is observed for excitation powers up to 4.35 $\mu$W indicating suppressed Auger recombination. }
\label{SFigFlatLifePLComp}
\end{figure}

In order to observe evidence of charge transfer, which we expect to be present in our WSe$_2$ monolayer on Au sample \cite{Zhang2019a}, we recorded the neutral exciton emission from a flat part of the single layers on both the SiO$_2$ and Au substrates under pulsed excitation at 638 nm focused to a 2.2 $\mu$m spot, displayed in Figure \ref{SFigFlatLifePLComp}(a). We observed an order of magnitude reduction in the PL from the Au surface compared to the SiO$_2$ substrate suggesting additional non-radiative decay channels have been introduced. Following this, we observed further evidence of a charge-transfer process in power dependent PL decay measurements.

Exciton-exciton annihilation processes have been previously observed to dominate room temperature neutral exciton emission from WSe$_2$ monolayers \cite{Mouri2014a} leading to a quick exponential decay of the emission soon after excitation (<100 ps) followed by radiative and further non-radiative decay processes resulting in a longer exponential decay ($\approx$ 1 ns). This results in a biexponential emission decay curve. We have measured WSe$_2$ monolayer neutral exciton emission decay using 90 ps pulsed excitation at a repetition rate of 80 MHz for a range of powers from 39 nW to 40 $\mu$W from the SiO$_2$ and Au substrates, plotted in Figures \ref{SFigFlatLifePLComp}(a) and (b) respectively. For powers below 125 nW we observe an exponential decay curve from the monolayers on both substrates. At higher powers, a second exponential component of the emission decay is observed to emerge for the WSe$_2$ monolayer on the SiO$_2$ substrate. This then quickly begins to dominate the entire decay at powers as high as several tens of $\mu$W as expected from exciton-exciton annihilation \cite{Mouri2014a}. In contrast to this, the monolayer on the Au film does not exhibit a non-negligible second exponential component in its decay trace until powers above 4.35 $\mu$W. We attribute this to the fast charge-transfer expected to occur for a WSe$_2$ monolayer deposited onto an Au substrate \cite{Zhang2019a} which depletes the exciton population and leads to a suppressed exciton-exciton annihilation rate \cite{Mouri2014a}. The larger diffusion lengths of dark excitons (1.5 $\mu$m \cite{Cadiz2018}), believed to feed the state responsible for single photon emission in monolayer WSe$_2$ \cite{Sortino2021}, compared to our 190 nm WS$_2$ NAs on the Au substrate, atop which the strain is most likely expected to form quantum emitters, suggests exciton-exciton annihilation may also be partially suppressed near the single photon emitters.

\FloatBarrier
\pagebreak

\begin{center}
\Large
\textbf{Supplementary Note 4: Heating effects due to absorption from Au substrate}
\end{center}

\begin{figure}[hb!]
	\centering
  \includegraphics[width=\linewidth]{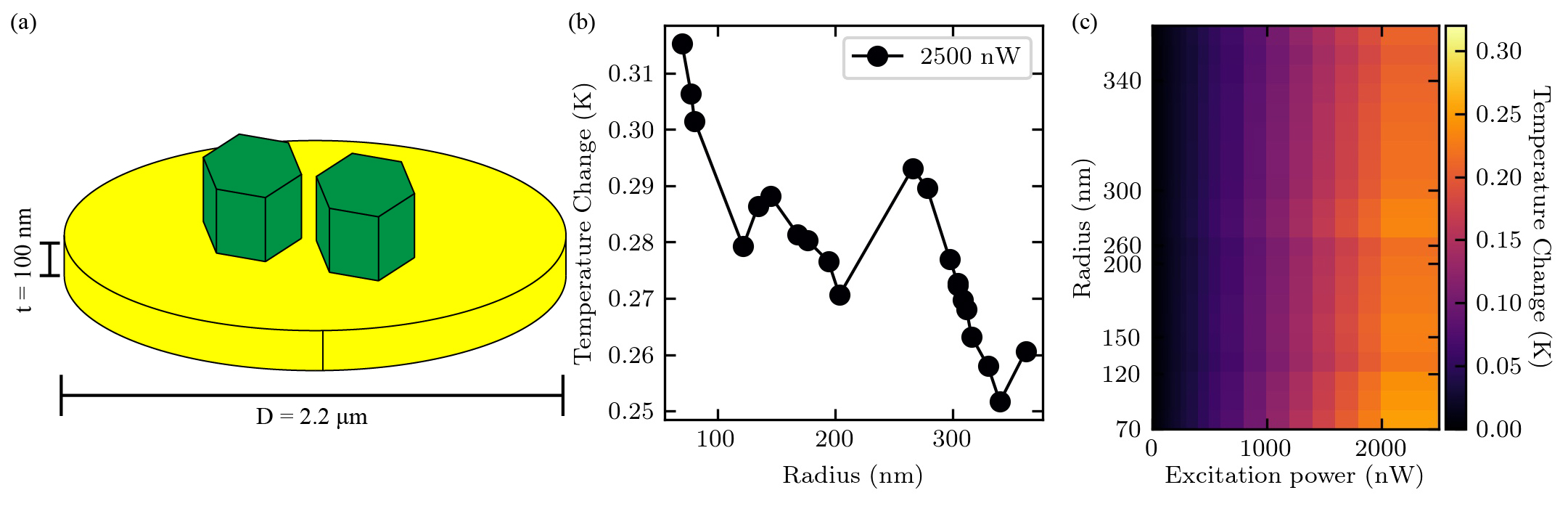}
  \caption{\textbf{Calculated change in temperature of the WS$_2$ NA on a Au substrate sample due to absorption in the metallic film.} 
	\textbf{(a)} Schematic illustration of a WS$_2$ dimer NA on a Au disk used for calculating the change in temperature of the sample under 638 nm pulsed (5 MHz) excitation. NA radii vary over the experimentally achieved range while the height is 190 nm and the dimer gap is 400-500 nm as recorded for the fabricated WS$_2$ NAs on the Au substrate.
	\textbf{(b)} Plot of the calculated temperature change in the sample due to absorption of the excitation light by the Au film for WS$_2$ NAs with the experimental range of radii at a maximum excitation power of 2.5 $\mu$W used in the SPE study.
	\textbf{(c)} Plot of the expected temperature change of the sample due to absorption of the excitation light by the Au film for WS$_2$ NAs with the experimental range of radii at the full range of excitation powers used in the SPE study. }
\label{SFigHeating}
\end{figure}

As WSe$_2$ single photon emitter intensities are strongly suppressed at temperatures above 30 K \cite{Luo2019}, we considered that any heating effects due to the absorption of the Au substrate \cite{Johnson1972}, may lead to a reduction in the peak intensities of the SPEs. We do not expect any heating due to absorption in the WS$_2$ nanoantennas as the excitation wavelength of 637 nm is below the exciton absorption of the material (625 nm) \cite{Zotev2023}. In order to understand how the metal absorption of the exciting pulsed laser light may locally increase the temperature of the sample, we consider a simple geometry for which the calculation is easier. This consists of a WS$_2$ dimer nanoantenna fabricated on an Au disk of thickness identical to the film in our experiments (100 nm) and a diameter (2.2 $\mu$m) equal to the excitation spot size (see Figure \ref{SFigHeating}(a)). As this encompasses the entire region which may absorb the illumination and the thermal contact with the rest of the film is expected to reduce the temperature surrounding the nanoantenna further, our simplified geometry will provide an upper bound to the heating of the fabricated sample. The first step is to decide whether a pulsed excitation model is required or whether a continuous wave (CW) model may be sufficient. A simple parameter $\xi = f\tau_d$, which is a multiplication of the repetition rate of the laser ($f$) and the cooling time of the Au disk ($\tau_d$), can be used to gauge which model is necessary. At values of $\xi << 1$, the heat in the Au disk would dissipate in between consecutive excitation pulses and a pulsed model would be required to extract the maximum temperature of the sample geometry. At values of $\xi \approx 1$ and above, the sample heat would no longer be dissipated in between excitation pulses and a CW model becomes sufficient to calculate the maximum local temperature of the sample. The cooling time of the Au disk ($\tau_d$) can be reformulated so that the $\xi$ parameter can be written as follows \cite{Baffou2011}:

\begin{equation}
\xi = \frac{f}{\tau_d} = fR^{2}_{eq}\frac{\rho_{Au}c_{Au}}{\kappa_{He}},
\label{eq:1}
\end{equation}

\noindent
where $\rho_{Au}$ is the mass density of Au at a temperature of 4K, $c_{Au}$ is the heat capacity of Au and $\kappa_{He}$ is the thermal conductivity of low temperature gaseous Helium, a small amount of which is used for thermal exchange in order to cool the sample to liquid helium temperatures despite being pumped to vacuum. Using known literature values for each of these numbers, we calculate a parameter $\xi = 23.88$ which is far above unity and thus a CW heating model is sufficient to calculate the maximum temperature increase of the sample geometry. The local increase in temperature of our sample ($\Delta T_{CW}$) can then be written as follows \cite{Baffou2010}:

\begin{equation}
\Delta T_{CW} = \frac{I\sigma_{abs}}{4\pi R_{eq}\beta\kappa_{He}},
\label{eq:2}
\end{equation}

\noindent
where $I$ is the illumination intensity, $\sigma_{abs}$ corresponds to the absorption cross section, $R_{eq}$ is the radius of an Au sphere with the same volume as the disk and $\beta$ corresponds to a dimensionless thermal-capacitance coefficient for an Au disk \cite{Baffou2010}. As this same model is used for the calculation of laser heating for an Au sphere, the latter coefficient, which is a function of the aspect ratio of the Au disk, is used to account for the fact that our calculation is not performed for a sphere, which is the shape with the lowest heat dissipation. The $\beta$ coefficient for a disk can be written as follows \cite{Baffou2010}:

\begin{equation}
\beta = e^{0.04 - 0.0124ln(\frac{D}{d}) + 0.0677ln^{2}(\frac{D}{d}) - 0.00457ln^{3}(\frac{D}{D})},
\label{eq:3}
\end{equation}

\noindent
where $D$ is the diameter of the disk and $d$ is its thickness. We thus substituted $\beta$, a literature value for $\kappa_{He}$, a calculation of $R_{eq}$ from the diameter and thickness of the gold disk, the maximum excitation intensity used in our experiments (2500 nW/$\mu$m) and numerical simulations of the absorption cross section in the metal film due to the nanoantenna mode confinement into equation \ref{eq:2}. The results of this calculation are plotted in Figure \ref{SFigHeating}(b) for the range of nanoantenna radii we have fabricated on the Au substrate. Similar to the electric field intensities seen in Figure 1(f) of the main text, as a result of the nanoantenna mode confinement, the highest temperature increase (0.315 K) is expected at the smallest NA radius. We have also plotted, in Figure \ref{SFigHeating}(c), the expected maximum local temperature change in the sample for the full range of excitation powers used in our measurements and the range of experimentally accessible nanoantenna radii. As this model only provides an estimate of the temperature increase for a simple nanoantenna on gold disk geometry, the increased thermal contact with the rest of the Au film, which is not absorbing in our experiments, is expected to lead to even smaller temperature increases in the experimentally studied geometry. These calculations thus suggests that the temperature increase as a result of the absorption in the metal film only negligibly contributes to the intensity reduction of the SPEs forming on NAs on the Au film as opposed to the SiO$_2$ substrate as plotted in Figure 2(c) of the main text.

\FloatBarrier

\begin{center}
\Large
\textbf{Supplementary Note 5: Experimental setup efficiency}
\end{center}

The experimental setup collection efficiency used in the SPE experiments was estimated by measuring the losses of a 725 nm CW laser passing through this setup. The transmission values through the individual components of the experimental setup are recorded in the table below. The values for transmission through the linear polarizer, the spectrometer and the CCD efficiency are taken from the relevant datasheets. The total experimental setup collection efficiency was estimated to be 0.56$\%$.

\begin{table}[H]
\centering
\begin{tabular}{|c|c|}
\hline
\textit{\textbf{Transmission}} & \textit{\textbf{Component}} \\
\hline
\hline
50$\%$ & Optical components \\
\hline
78$\%$ & Linear polarizer \\
\hline
2$\%$ & Single mode fibre coupling \\
\hline
90$\%$ & Spectrometer in-coupling \\
\hline
3$\times$ (97$\%$) = 91$\%$ & Spectrometer mirrors (x3) \\
\hline
90$\%$ & Grating efficiency \\
\hline
98$\%$ & CCD quantum efficiency \\
\hline 
\textbf{0.56$\%$} & \textbf{Total setup efficiency} \\
\hline
\end{tabular}
\caption{\textbf{Experimental setup collection efficiency.} The collection efficiency was estimated by measuring the losses of a 725 nm CW laser through our experimental setup. Transmission through the polarizer, spectrometer and the CCD efficiency are extracted from the relevant datasheets.}
\label{tab1}
\end{table}

\FloatBarrier

\begin{center}
\Large
\textbf{Supplementary Note 6: Optimized WS$_2$ NAs on Au substrate with dielectric spacer for a dimer gap achievable with only nano-fabrication}
\end{center}

\begin{figure}[ht!]
	\centering
  \includegraphics[width=\linewidth]{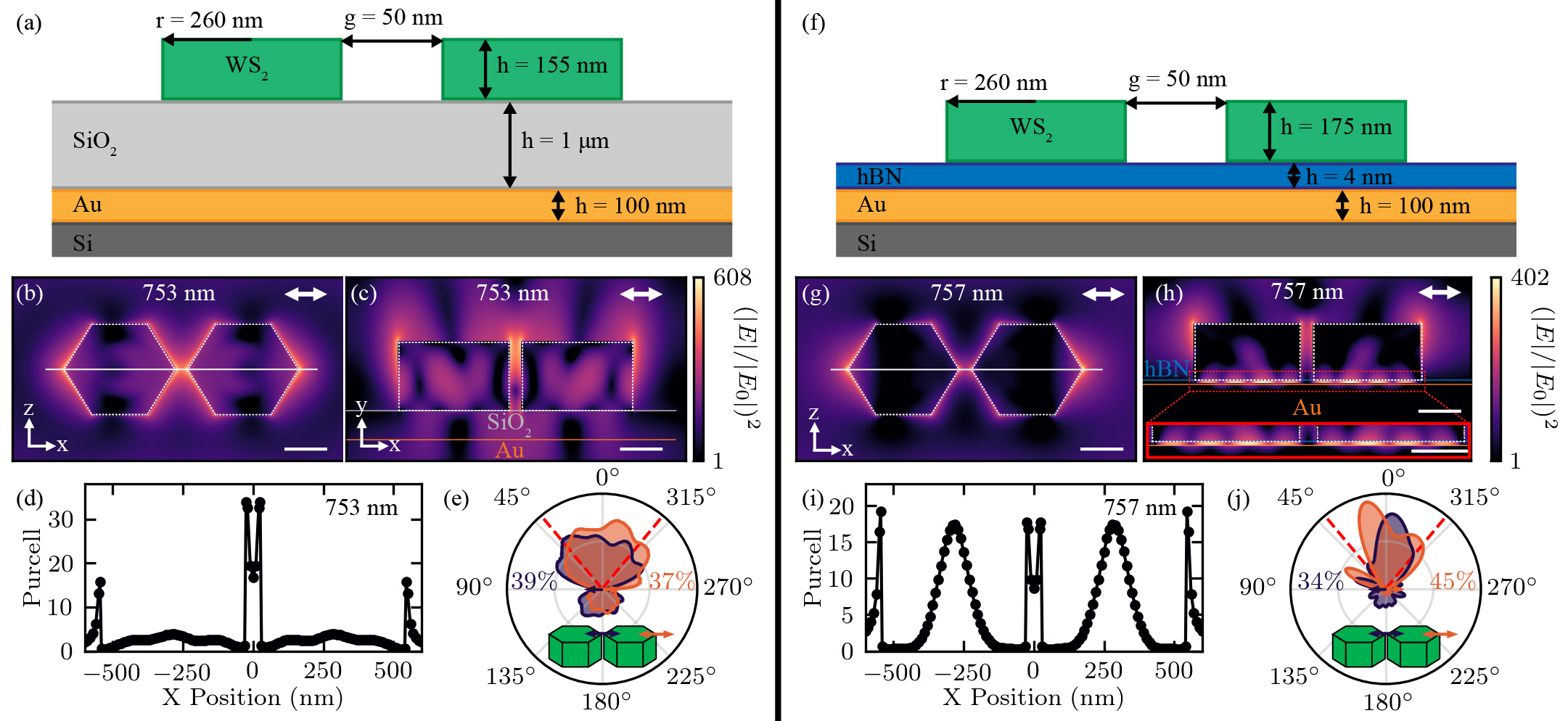}
  \caption{\textbf{Photonic enhancement simulations of optimized WS$_2$ NA designs on a Au substrate with a SiO$_2$ or hBN spacer for a dimer gap of 50 nm.}
	\textbf{(a)} and \textbf{(f)} Schematic illustrations of WS$_2$ NAs on a Au substrate with a SiO$_2$ and hBN spacer for optimum fluorescence enhancement of WSe$_2$ monolayer SPEs and a dimer gap of 50 nm achievable with standard nanofabrication.
	\textbf{(b)},\textbf{(c)} and \textbf{(g)},\textbf{(h)} Simulated electric field intensity profiles at the top surface and as a vertical cut through the middle of the WS$_2$ NAs for optimized designs with a SiO$_2$ and hBN spacer at the photonic resonance wavelength of 753 nm and 757 nm respectively. Electric field hotspots are still observed within the dimer gap and at the outer top vertices of the NAs. Scale bars = 200 nm.
	\textbf{(d)} and \textbf{(i)} Line plots of the Purcell factor at different positions at the top surface of the WS$_2$ NA along the dimer axis for the SiO$_2$ and hBN spacer designs respectively. Larger enhancements are observed within the dimer gap and at the outer top vertices of the NA, where SPEs are expected to form, as expected from the electric field hotspots in \textbf{(b)},\textbf{(c)},\textbf{(g)} and \textbf{(h)}.
	\textbf{(e)} and \textbf{(j)} Radiation patterns for SPE positions within the dimer gap (dark blue) and at the outer top vertices (orange) for the SiO$_2$ and hBN spacer designs respectively. Emission is directed mainly towards collection optics with the cone defined by the experimental numerical aperture in this study indicated by the red dashed lines. The numbers to the sides indicate the collection efficiency, with a numerical aperture of 0.64, for an emitter forming in the dimer gap (dark blue) and at the outer top vertices of the NA. Lower insets indicate the position of the SPE dipole relative to the WS$_2$ NAs.}
\label{SFigOptimizedG50}
\end{figure}

We performed a similar optimization procedure for the SiO$_2$ and hBN spacer placed between WS$_2$ hexagonal dimer nanoantennas and a metallic mirror for maximum fluorescence enhancements from WSe$_2$ monolayer SPEs. We did, however, constrain the dimer gap distance to 50 nm which can be achieved with standard nanofabrication techniques and does not require post-fabrication atomic force microscopy repositioning \cite{Zotev2022}. This optimization procedure yielded the same radius, nanoantenna height and spacer thicknesses as for a gap distance of 10 nm as shown in Figure \ref{SFigOptimizedG50}(a) and (f). The electric field intensities surrounding the optimized geometries at the same resonance wavelengths (756 nm and 752 nm) are plotted in Figure \ref{SFigOptimizedG50}(b)-(c) and (g)-(h) for the SiO$-2$ and hBN spacer designs respectively. The midgap hostpots yield electric field intensities of 608 and 142 while the outer vertex positions result in values of 461 and 402 for the SiO$_2$ and hBN spacers respectively. These values are not far lower than for the 10 nm gap distance design.

As we expect the Purcell enhancement to be highest at the inner and outer vertices of the nanoantennas, we only performed these simulations at positions along the dimer axis as shown by the white line in Figure \ref{SFigOptimizedG50}(b) and (g). The Purcell factors are plotted in Figure \ref{SFigOptimizedG50}(d) and (j). As the gap distance is increased to 50 nm, the Purcell factors expected for an emitter forming near an inner vertex of the nanoantenna has exponentially dropped to values of 34 and 18 for the SiO$_2$ and hbN spacer designs respectively. However, the Purcell enhancements near the outer vertices, where emitter formation is far more likely, are not much lower than for the designs presented in the main text reaching values of 16 and 19 for the SiO$_2$ and hBN spacer designs respectively. 

Lastly, the radiation patterns for emitters forming atop these nanoantenna designs on a metallic mirror substrate with a dielectric spacer also result in more of the light being directed toward the collection optics yielding similarly high collection efficiencies within our experimental numerical aperture. For an emitter forming in the dimer gap, the collection efficiencies are 39\% and 34\% for the SiO$_2$ and hBN spacer designs respectively. For an emitter forming at the outer vertices of the nanoantennas the collection efficiencies are 37\% and 45\% for the SiO$_2$ and hBN spacer designs respectively.

For an SPE position within the dimer gap, these 50 nm gap distance optimized designs yield much lower electric field intensities and Purcell factors leading to much lower expected fluorescence enhancement factors under resonant excitation within our experimental setup of more than 47000 and 5000 for the SiO$_2$ and hBN spacers respectively when compared to an emitter in vacuum. Upon comparison with an emitter forming on a flat 300 nm SiO$_2$/Si substrate, we calculate maximum fluorescence enhancement factors of more than 313000 and 33700 for the SiO$_2$ and hBN spacers respectively. However, positions atop the outer vertices of the NAs, where SPE formation is far more likely due to the high strain, yield only slightly reduced values of the electric field intensity and Purcell factor with similar collection efficiencies. Therefore, this optimized design with a larger dimer gap is expected to induce similarly high fluorescence enhancement factors, compared to vacuum, of 15910 and 20038 for the SiO$_2$ and hBN spacers respectively. When compared to an emitter forming on a flat 300 nm SiO$_2$/Si substrate, the enhancement factors reach 106070 and 133586 for the SiO$_2$ and hBN spacer designs respectively.

\FloatBarrier

\bibliographystyle{unsrt}
\bibliography{./library}